\documentclass[aps,twocolumn,pr*,amsfonts,showpacs,superscriptaddress,longbibliography]{revtex4-2}
    \DeclareUnicodeCharacter{0393}{$\Gamma$}
\usepackage{verbatim,epsfig,amsmath,amssymb,bm,epsf,graphicx,psfrag,bbold,amsthm,amsfonts}
\usepackage[bottom]{footmisc}
\usepackage{hyperref}
\usepackage{cleveref} 
\usepackage{enumitem}
\usepackage{framed}
\usepackage{mathrsfs}
\usepackage{esint}
\setlist{nosep}
\usepackage{placeins}
\usepackage[all]{xy}
\usepackage{color}
\usepackage[utf8]{inputenc}
\usepackage{float}
\usepackage{natbib}
\usepackage{tikz-cd}
\usepackage{verbatim}
\usepackage{leftidx}
\usepackage{physics}
\usepackage{ulem}

\usepackage{pifont}
\theoremstyle{plain}

\theoremstyle{definition}

\theoremstyle{remark}

\begin{document}

\title{Ferromagnetic superconductivity with excitonic Cooper pairs:\\ Application to $\Gamma$-valley twisted semiconductors}

\author{Daniele Guerci}
\affiliation{Department of Physics, Massachusetts Institute of Technology, Cambridge, Massachusetts 02139, USA}
\author{Liang Fu}
\affiliation{Department of Physics, Massachusetts Institute of Technology, Cambridge, Massachusetts 02139, USA}

\date{\today}

\begin{abstract}

We present a theory of ferromagnetic superconductivity that emerges upon doping a correlated ferromagnetic insulator through the condensation of excitonic Cooper pairs, which are charge-$2e$ bosonic quasiparticles made of Cooper pairs strongly hybridized with excitons. 
By solving a model of spin-polarized electrons using the strong-coupling expansion to the second order, we demonstrate the emergence of  excitonic Cooper pairs from electron-hole fluctuations upon doping a strongly correlated insulator. We characterize their binding energy, effective mass, and the resulting superconducting transition temperature. 
We propose possible realization of spin-polarized superconductivity in twisted semiconductors with honeycomb moir\'e superlattice.

\end{abstract}

\maketitle

{\it Introduction---} The search of high-temperature superconductivity driven by electron repulsion has long fascinated researchers due to potential technological applications and fundamental scientific interest. Since the pioneering work of Kohn and Luttinger \cite{kohnluttingerSC_prl}, superconductivity has been theoretically obtained from repulsive interaction in Fermi liquids, where the effective attraction arises from the oscillatory component of screened interaction. As the Kohn-Luttinger-type theories are based on interaction expansion~\cite{chubukov_1993,gonzalez_2008,Raghu2010,ErezKL_2021,guinea_2022,Schrade2024}, it only yields weak-coupling superconductivity, whose transition temperature $T_c$ is orders of magnitude smaller than the Fermi energy and coherence length far exceeds interparticle distance.  

Recently, a novel mechanism for superconductivity from repulsive interaction has been introduced for multiband systems \cite{Slagle_prb,Cr_pel_2021}. For simple models of correlated band insulators, it has been shown rigorously that an effective attraction between doped electrons can arise from {\it interband} charge fluctuations. These fluctuations  mediating superconductivity are associated with the ``vibrations'' of the valence electrons (i.e., excitons) \cite{Cr_pel_2022}, as opposed to the ion lattice vibrations (i.e., phonons) in conventional superconductors. Possible applications of this electronic pairing mechanism have been discussed for various models and materials \cite{Cr_pel_Cea_2022,chou_2022,PhysRevResearch.5.L012009,homeier2023feshbachhypothesishightcsuperconductivity, crepel2023prl,YangSC_2024,Milczewski_2024,Zerba_2024,takahashi_2025}. 

In this work, we present a theory of ferromagnetic superconductivity that arises from doping a strongly correlated ferromagnetic insulator. This unconventional superconducting state is spontaneously, fully spin-polarized and features tightly bound electron pairs dressed with excitons. 
By solving a minimal model of strongly interacting electrons on the honeycomb lattice, we show explicitly that two important energy scales for superconductivity---the pairing gap and the superfluid stiffness---are both controlled by the interaction strength in our system. The maximum superconducting transition temperature $T_c$ reaches a significant fraction of the bandwidth.   

\begin{figure}
    \centering
        \includegraphics[width=\linewidth]{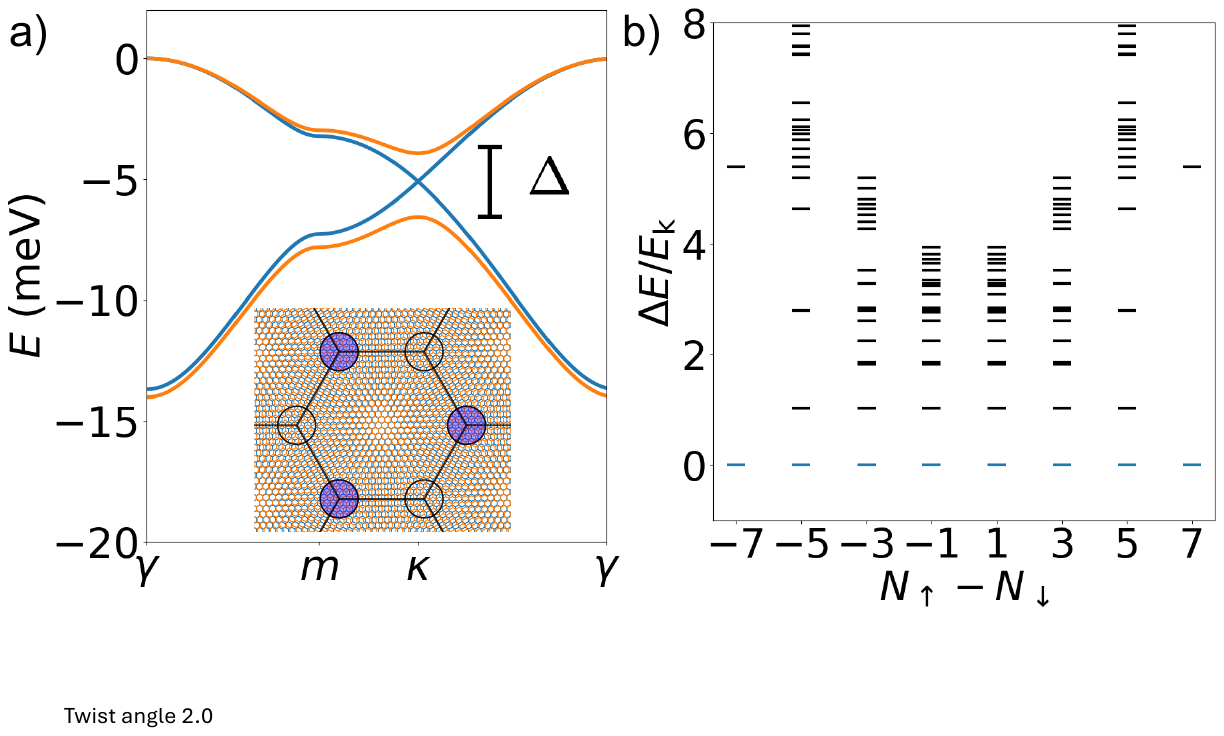}
    \caption{\textbf{$\Gamma$-valley twisted semiconductors}: Panel a) shows the low-energy bands, with a tunable gap $\Delta$ controlled by $D$. Blue and orange denote $D=0$ and $D=60$meV, respectively. The inset displays the moir\'e pattern with high-symmetry stackings (MM, MX, XM). Wannier orbitals localize at MX and XM, forming a honeycomb lattice. Panel b) shows the many-body spectrum in units of $E_{\rm k} =\hbar^2/(2ma^2)$ across different spin $S^z$ sectors for a $3 \times 3$ cluster, including the two topmost bands. The calculations are performed with $\epsilon = 10$, $d_{\rm sc} = 5$nm and $\theta = 2^\circ$. }
    \label{fig:model_exciton}
\end{figure}

Our theory is motivated by  $\Gamma$-valley twisted transition metal dichacogenides ($t$TMD)~\cite{yangzhang_prb,Angeli_2021,Xian_2021,Haining_2023}, where Wannier orbitals are centered at the MX and XM moir\'e sites forming a honeycomb lattice (inset of Fig.~\ref{fig:model_exciton}a). At small twist angle, the low-energy moir\'e bands exhibit Dirac points similar to graphene, but has a very narrow bandwidth (Fig.~\ref{fig:model_exciton}a) suitable for strongly correlated phenomena. Indeed, a recent experiment \cite{ma2024relativisticmotttransitionstrongly} has observed correlated insulators at the filling of $\nu=1$ hole per unit cell. 

{\it Ferromagnetism---} $\Gamma$-valley moiré semiconductors have negligible spin-orbit coupling~\cite{Korm_nyos_2015}, which leads to spin SU(2) symmetry~\cite{yangzhang_prb,Angeli_2021}.
By exact diagonalization of the interacting continuum model for $\Gamma$-valley $t$TMDs, detailed in the Supplementary Material (SM)~\cite{supplementary}, we find robust ferromagnetism over a wide range of twist angles and interaction strengths, both at filling $\nu=1$ and under finite hole doping. Notably, these ferromagnetic ground states are fully spin polarized, possessing $(2S+1)$-fold degeneracy with $S=N/2$ ($N$ is the total number of spin-$1/2$ electrons).  
This behavior is especially pronounced near $\nu = 1$, as illustrated in Fig.~\ref{fig:model_exciton}b), which shows the many-body spectrum across different total spin $S^z$ sectors for $\nu=7/9$. 
In addition, we observe $(2S-1)$-fold degeneracy in low-lying excited states, consistent with one-magnon excitation. 
We also note that ferromagnetism extends to $\nu=1$, where strong electron interaction induces a correlated ferromagnetic insulator with broken sublattice symmetry. 

{\it Extended Hubbard model on honeycomb lattice---} Building on our continuum model results, we study a minimal model of spin-polarized fermions on a honeycomb lattice, incorporating the shortest-range non-trivial repulsive interactions:
\begin{equation}\label{tVmodel}
    H = -t \sum_{\langle r,r' \rangle} f^\dagger_r f_{r'} + V \sum_{\langle r,r' \rangle} n_r n_{r'}
    +\Delta N_B, 
\end{equation}
where $\langle r,r' \rangle$ denotes nearest-neighbor (n.n.) sites on the honeycomb lattice and $N_{A,B}$ the total number of particles on $A$ or $B$ sublattice, which corresponds to MX and XM moir\'e sites respectively. 
$\Delta$ represents the potential difference between the two sublattices, which is induced by an applied displacement field $D$, as shown in the SM~\cite{supplementary}.

\begin{figure}
    \centering
    \includegraphics[width=0.8\linewidth]{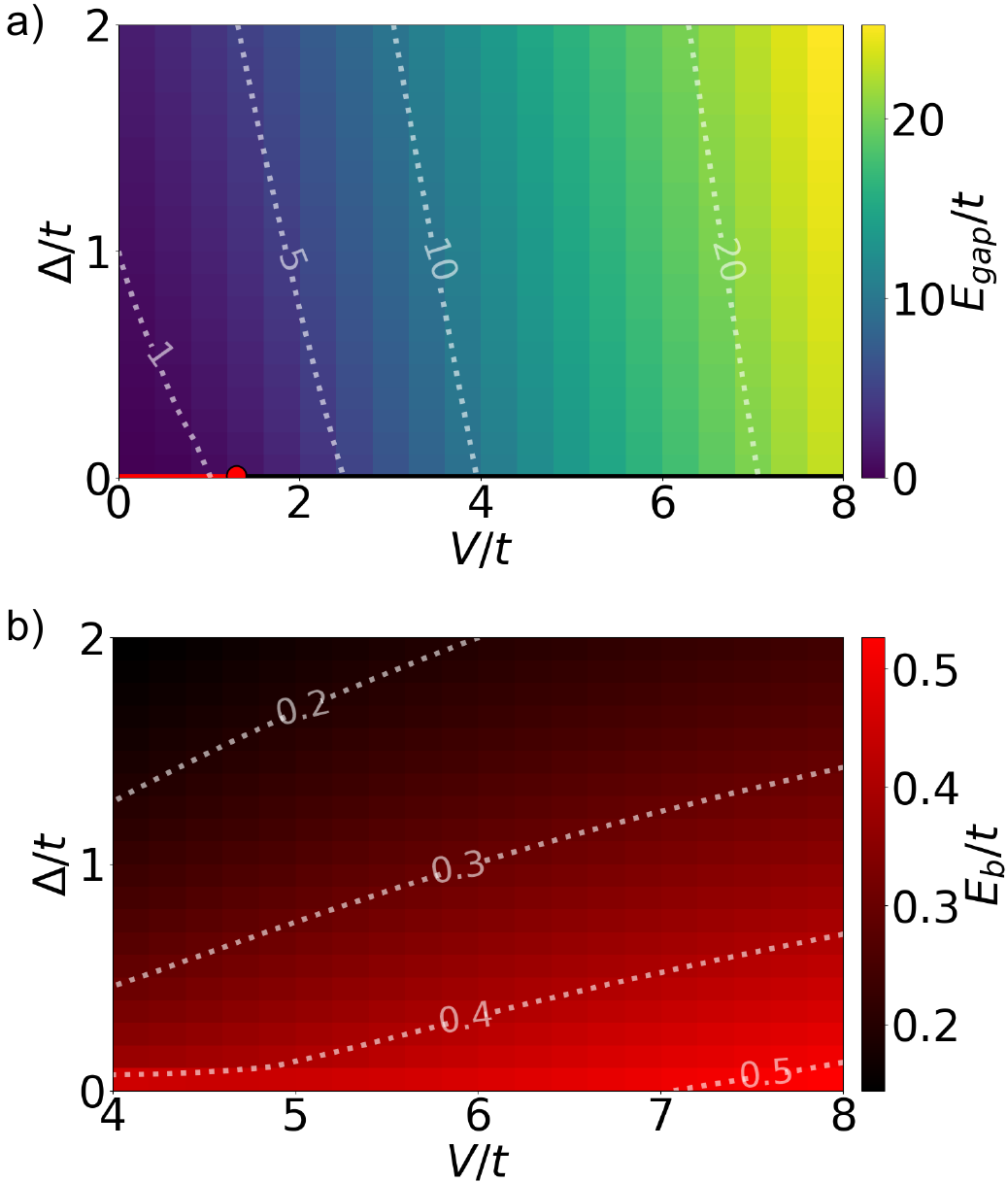}
   \vspace{-0.3cm}
    \caption{\textbf{Phase diagram at $\nu=1$ and finite doping}: Panel a) shows the charge gap at $\nu=1$. 
    The red dot at $\Delta=0$ marks the critical interaction $V_c=1.3t$, separating the Dirac semimetal from the sublattice-polarized insulator. Panel b) shows the binding energy of excitonic Cooper pairs. {ED is performed on a 24-site lattice with periodic boundary conditions}.}
    \label{fig:fig2}
\end{figure}

For large $V/t$, the ground state at $\nu=1$ is a gapped insulator;  the charge gap obtained from our exact diagonalization (ED) calculation is shown in Fig.~\ref{fig:fig2}a). Depending on the sign of $\Delta$, either $A$ or $B$ sites are preferentially occupied, while at $\Delta=0$, the system spontaneously breaks the sublattice symmetry~\cite{gross_1974,Semenoff_1984,Herbut_2009,Herbut_2006,Juricic_2009} at $V> V_c =1.3 t$~\cite{supplementary}, consistent with previous studies~\cite{Wang_2014,Troyer_2016,capponi2016phase,Huffman_2017,Schuler_2021}. 
In this work we will focus on the strongly interacting regime $V/t \geq 5$ where the correlation length is short, which justifies our strong-coupling expansion in $t/V$ and mitigates finite-size effect in our ED study. 

{\it Excitonic Cooper pair---} In order to find the ground state at small doping $\nu=1 + \delta$, we analyze the energy cost of various charge-$e$, $2e$ and $4e$ excitations of the $\nu=1$ correlated insulator. The model is particle-hole symmetric, showing identical behavior for electron doping ($\delta>0$) and hole doping ($\delta<0$). 
Before presenting the full theory, we summarize our first main finding in Fig.~\ref{fig:fig2}b): the binding energy $E_b$ of two doped particles, which signals the emergence of excitonic Cooper pairs. 

To understand the origin of pairing from repulsive interaction in our model, let us first consider charge excitations in the infinite coupling limit $V\rightarrow \infty$. 
Here, the ground state at $\nu=1$ is fully sublattice polarized and quantum fluctuation is completely suppressed, because any hopping process entails interaction energy cost $V$.        
Assuming that $A$ sites are occupied at $\nu=1$, adding a single particle to a $B$ site costs energy 
$E_{1\rm e} = \Delta+ 3V$. This charge-$e$ particle is also ``frozen'' because moving it costs additional interaction energy.   
\begin{figure}
    \centering
    \includegraphics[width=0.7\linewidth]{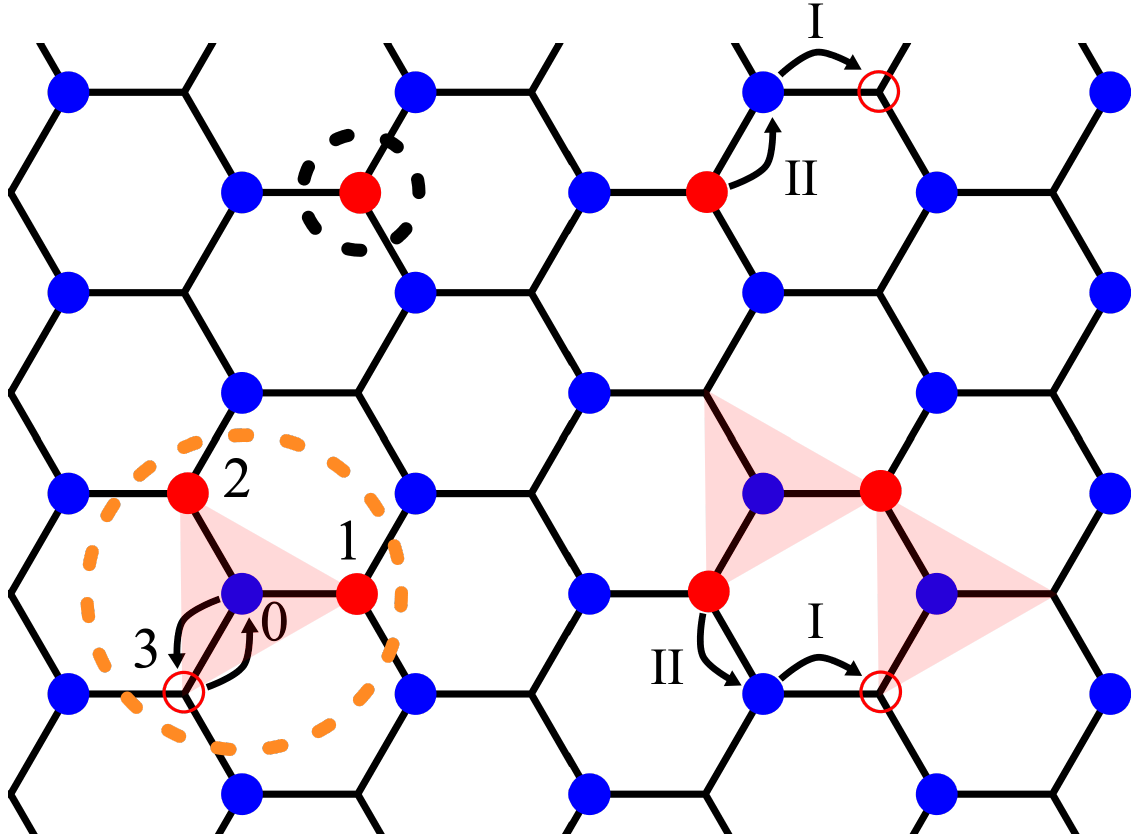}
    \vspace{-0.3cm}
    \caption{\textbf{Excitonic Cooper pair and charge carrier motion}: 
    Left: Charge-$e$ and $2e$ excitations, highlighted by dashed circles, correspond to an excitonic Cooper pair (orange) and a fermion (black). Right: Leading-order processes contributing to quasiparticle mass: (I) polaron formation and (II) its recombination, inducing center-of-mass motion.}
    \label{fig:fig3}
\end{figure}

On the other hand, consider a pair of particles added to two neighboring $B$ sites, denoted as $1$ and $2$ in Fig.~\ref{fig:fig3}. 
This configuration is connected by the hopping term $f^\dagger_{3} f_0$ to a ``trimer'' configuration, with a cluster of three particles on $B$ sites $1,2,3$ surrounding an empty $A$ site $0$~\cite{Slagle_prb}. 
Importantly, the trimer configuration costs the same interaction energy $6V$ as the initial configuration. Therefore, even at $V\rightarrow \infty$, quantum hopping $t$ leads to a linear superposition between a localized pair of particles and a trimer---the latter is dressed by a charge-transfer exciton ($f^\dagger_{3} f_0$) as shown in Fig.~\ref{fig:fig3}. Since coherent superposition lowers the energy, the resulting charge-$2e$ complex---which we call ``excitonic Cooper pair''---is lower in energy than two separate charge-$e$ particles.     

In the $V\rightarrow \infty$ limit, an isolated excitonic Cooper pair cannot move because any hopping term will only connect it to configurations that cost additional interaction energy $V$. This allows us to determine its binding energy per particle $E_b\equiv E_{1e}-E_{2e}/2$, where $E_{2e}$ is the charge-$2e$ excitation energy, exactly by solving our model~\eqref{tVmodel} on a four-site cluster, yielding: 
\begin{equation}\label{2e_binding_energy}
    E_b = \sqrt{\Delta^2/16+3t^2/4}-\Delta/4.
\end{equation}
The many-body wavefunction of the system having a single excitonic Cooper pair centered at $A$ site $r$ is: 
\begin{eqnarray}\label{single_pair}
    \ket{\Phi_2(r)} &=& \left( \frac{\alpha}{\sqrt{3}} \sum_{j=1}^{3} f^\dagger_{r'_j} f^\dagger_{r'_{j+1}}  
    + \sqrt{1-\alpha^2} f^\dagger_{r'_1} f^\dagger_{r'_2} f^\dagger_{r'_3} f_r \right) \ket{\Phi_0} \nonumber \\
    &\equiv& b^\dagger_r\ket{\Phi_0}, 
\end{eqnarray}
where $\ket{\Phi_0}=\prod_{r\in A}f^\dagger_r\ket{0}$ is the undoped ground state; $r'_j$ with $j=1,2,3$ denotes the three $B$ sites adjacent to $r$.    
The excitonic Cooper pair wavefunction resonates between a Cooper pair and a trimer, displayed in Fig.~\ref{fig:fig3}, with probability $\alpha^2$ and $1-\alpha^2$ respectively, where $\alpha$ depends on $\Delta/t$, $\alpha^2=1/2+\Delta/\sqrt{4\Delta^2+48t^2}$. 
The pair wavefunction belongs to the $A_2$ irrep of $D_3$, exhibiting $f$-wave symmetry~\cite{supplementary}.

Our strong-coupling result at $V \gg t$ complements the previous study in a different regime $\Delta \gg t$~\cite{Cr_pel_2021}. Importantly, our result shows that for large repulsive interaction,  the Cooper pair is strongly hybridized with the exciton at small $\Delta$, resulting in a large pair binding energy which reaches the maximum value $E_b|_{\Delta=0}=\sqrt{3}t/2$ at $\Delta=0$. As $\Delta$ increases,  the hybridization with the exciton is reduced; the pair binding energy decreases monotonously and becomes vanishingly small $E_b \approx 3t^2/\Delta$ in the limit $\Delta\gg t$ in agreement with Ref.~\cite{Cr_pel_2021}. 
In the rest of this work we focus on the regime of large $V/t$ and small $\Delta/t$, where the exciton binds two doped particles tightly together.     

Next, we perform a strong-coupling expansion in the small parameter $t/V$ to study the regime of large but finite interaction strength.  The strong coupling expansion is performed by organizing the Hilbert space into sectors having different numbers ($M$) of n.n. occupied sites. The Hamiltonian, when decomposed into these sectors, consists of a block diagonal term and an off-diagonal term, given by $H=H_{0}+H'$. 
$H_0$ is expressed as $H_{0}=\sum_{M}H_{M}$: 
\begin{equation}\label{H_unperturbed}
    H_{M}=-t\mathbb P_M\sum_{\langle r,r'\rangle}f^\dagger_rf_{r'}\mathbb P_M + \Delta N_B +MV.
\end{equation}
Here, $\mathbb P_M$ is the projector onto the sector with $M$ n.n. occupied sites, and the term $MV$ is the interaction energy. The off-diagonal part, which couples sectors with different values of $M$, is given by  $H'=\sum_M\sum_{q\neq 0} T_{q,M}$:
\begin{equation}\label{tunneling_q}
    T_{q,M}=-t\mathbb P_{M+q}\sum_{\langle r,r'\rangle}f^\dagger_{r} f_{r'}\mathbb P_M,
\end{equation} 
where $T_{q,M}$ changes the number of n.n. occupied sites by $q=\pm1,\pm2$ in a sector with fixed $M$. 

In the absence of $H'$ (or $V\rightarrow \infty$), the ground states of $H_0$ with zero $(p=0)$, one $(p=1)$ and two $(p=2)$ doped particles have a fixed number of n.n. occupied sites $M=zp$, leading to ground state energies $E_p = z p V$ with $z=3$ coordination of the lattice. 
Note that in the presence of doped particles ($p=1,2$) the ground states of $H_0$ are extensively degenerate as discussed above. This degeneracy is lifted by virtual processes induced by 
$H'$, which couple the low-energy sector to high-energy sectors with $M=zp+q$ costing additional interaction energies $qV$. These virtual processes are accounted for using the Schrieffer-Wolff (SW) transformation~\cite{SW_1966,allan_1988,Bravyi_2011}, a unitary transformation that systematically eliminates the coupling between low- and high-energy sectors: ${\cal H}=e^{S} H e^{-S}$, where $S$ is anti-Hermitian. Importantly, the SW transformation can be carried out by a perturbative expansion in $t/V$: $S=S_1 + S_2 + ...$ with $S_j \sim (t/V)^j$.  

As detailed in the SM~\cite{supplementary}, we calculate $S$ up to the second order $(t/V)^2$, so that low- and high-energy sectors are decoupled in the transformed Hamiltonian $\cal H$  up to the order $(t/V)^2$.   
Projecting $\cal H$ onto the low-energy manifold with $M=zp$ yields the effective Hamiltonian of interest ${\cal H}^{(p)}$, with $p=0,1,2$ denoting the number of doped particles. ${\cal H}^{(p)}$ takes a particularly simple form at $\Delta=0$~\cite{supplementary}, 
\begin{widetext}
\begin{equation}\label{SWHamiltonian}
    \begin{split}
    &{\cal H}^{(p)}=H_{zp}-\sum_{q=1}^{2}\frac{T^\dagger_{q,zp}T_{q,zp}}{qV} +\sum_{q=1}^2\frac{T^\dagger_{q,zp}T_{0,zp+q}T_{q,zp}}{(qV)^2}-\frac{1}{2}\sum_{q=1}^2\frac{\{T^\dagger_{q,zp}T_{q,pz},T_{0,zp}\}}{(qV)^2},
    \end{split}
\end{equation}
\end{widetext}
where $\{\cdot,\cdot\}$ is the anticommutator. 

We now analyze the consequences of the above strong-coupling expansion for the undoped ground state and charge excitations. For the undoped case, the effective Hamiltonian ${\cal H}^{(0)}$ yields the correction to the ground state energy ${\delta E}_{0} =-N zt^2/(2V)$ up to $\mathcal O(t^4/V^3)$, where $N$ is the number of unit cells. 

For one doped charge ($p=1$), the perturbation due to $H'$ in the effective Hamiltonian lifts the degeneracy of charge-$e$ excitations and endows them with a energy-momentum dispersion. This is derived by projecting $\mathcal H^{(1)}$ in the degenerate manifold of unperturbed ground states $\ket{\Phi_1(r)}=f^\dagger_r\ket{\Phi_0}$. The resulting hopping Hamiltonian for a charge-$e$ quasiparticle takes the form: $\mathcal H^{(1)}=E_1+t_f\sum_{\langle r,r'\rangle\in B}f^\dagger_r f_{r'},$ where $t_f=t^2/V+\mathcal O(t^4/V^3)$ represents the hopping amplitude between adjacent $B$ sites illustrated in Fig.~\ref{fig:fig3}, and the constant energy term $E_{1}=zV+\delta E_1$ with $\delta E_1=\delta E_0-3zt^2/(2V)$ and $\delta E_0$ includes the energy correction arising from virtual processes. 
The corresponding dispersion relation is given by $\epsilon_f ({\bf k}) = E_1 + 2t_f\sum_j \cos({\bf k}\cdot {\bf a}_j)$~\cite{supplementary}. 

\begin{figure}
    \centering
    \includegraphics[width=\linewidth]{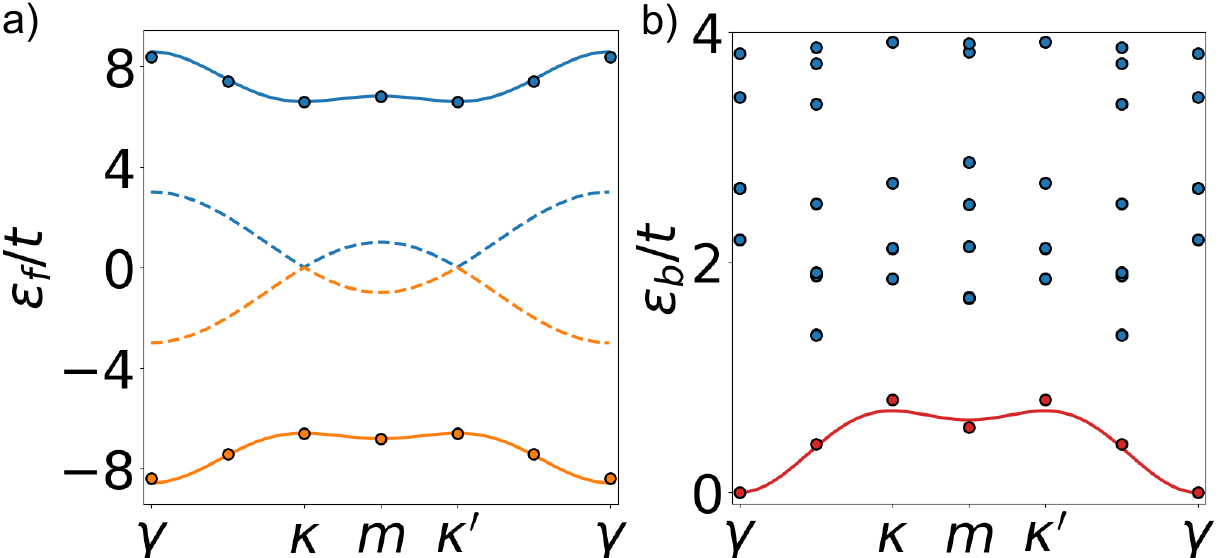}
    \vspace{-0.25cm}
    \caption{\textbf{Charge$-e$ and excitonic Cooper pair dispersion relation}: 
    Panel a) displays the $\pm e$ excitation spectra with (solid) and without (dashed) interactions.
    {Panel b) shows the charge-$2e$ quasiparticle dispersion relation}. 
    Dots: ED spectrum; solid lines: strong-coupling theory (no fit parameters).
    {We employed $(\Delta,V)/t=(0,5)$ on a 24-site cluster.}}
    \label{fig:dispersion}
\end{figure}

Fig.~\ref{fig:dispersion}a) shows the  band dispersion of charge-$1e$ excitations of the sublattice polarized insulator at $V/t=5$. Results obtained from ED (dots) and strong-coupling expansion (solid lines) are found to be in excellent agreement. For comparison and contrast, we also included the bare dispersion relation (dashed lines), which features Dirac cones. 

In the charge-$2e$ sector ($q=2$), excitonic Cooper pairs located at different $A$ sites $\ket{\Phi_2(r)}=b^\dagger_r\ket{\Phi_0}$~\eqref{single_pair}, are degenerate in the absence of $H'$ and form an orthonormal basis $\braket{\Phi_2(r)}{\Phi_2(r')}=\delta_{rr'}$.
After including perturbative corrections to second order in $t/V$, 
we obtain an effective Hamiltonian within this degenerate subspace which governs the hopping of excitonic Cooper pair:
\begin{equation}\label{bosonic_hamiltonian}
    \mathcal H^{(2)} = E_2 - t_b \sum_{\langle r,r'\rangle\in A} b^\dagger_r b_{r'}, 
\end{equation}
and the corresponding energy dispersion is given by $\epsilon_b({\bf q}) =E_2  -2 t_b \sum_{j=1}^{3}\cos({\bf q}\cdot {\bf a}_j)$, where $\bf q$ is the Cooper pair momentum.  
Here, $E_2=2\Delta +2zV-2E_b+ \delta E_2$ includes correction $\delta E_2=\delta E_0 -zt^2/V-5\sqrt{3}t^3/(4V^2)$ due to virtual processes, which will affect the binding energy to be discussed later.  
The hopping amplitude of excitonic Cooper pair $t_b$~\eqref{bosonic_hamiltonian}, for $\Delta=0$, is given by:
\begin{equation}\label{hopping_boson}\begin{split}
    t_b=\frac{t^2}{6V}+\frac{\sqrt{3}t^3}{2V^2}+\mathcal O\left(\frac{t^4}{V^3}\right). 
    \end{split}\end{equation}
Here, the leading order contribution $\sim t^2/V$ originates from the second-order particle hopping process illustrated in Fig.~\ref{fig:fig3}. 
Eq.~\eqref{hopping_boson} also includes the next leading order contribution $\sim t^3/V^2$, which originates from various third-order hopping processes as detailed in the SM~\cite{supplementary}.

Fig.~\ref{fig:dispersion}b) shows the energy dispersion of charge-$2e$ excitations. Results obtained from ED (dots) and our analytical expression (solid line) with $t_b$ given in Eq.~\eqref{hopping_boson} are found to be in excellent agreement. While the charge-$e$ fermion dispersion has degenerate minima at $K, K'$, the charge-$2e$ boson dispersion has the minimum at $\Gamma$, i.e., ${\bf q}=0$.  
Comparing the energy difference between the ground states of our system doped with one and two particles, we determine the binding energy up to order $\mathcal O(t^3/V^2)$, which for $\Delta=0$, is given by:
\begin{equation}
    E_b=\frac{\sqrt{3}}{2}t-\frac{3t^2}{V}+\frac{5\sqrt{3}t^3}{8V^2}+\mathcal O\left(\frac{t^4}{V^3}\right).
\end{equation}
Compared to $V=\infty$, the binding energy decreases monotonously as $V$ is reduced, 
but remains large $E_b \approx 0.31 t$ at $V/t=5$.   

We emphasize that all analytical results including binding energy and charge-$2e$ dispersion are obtained from strong-coupling expansion to second order in $t/V$ without any adjustable parameter. It is remarkable that analytical and ED results are in excellent agreement up to $t/V=0.2$. We further extend analytical calculations to include finite $\Delta$ in the SM~\cite{supplementary}.  

\begin{figure}
    \centering
    \includegraphics[width=\linewidth]{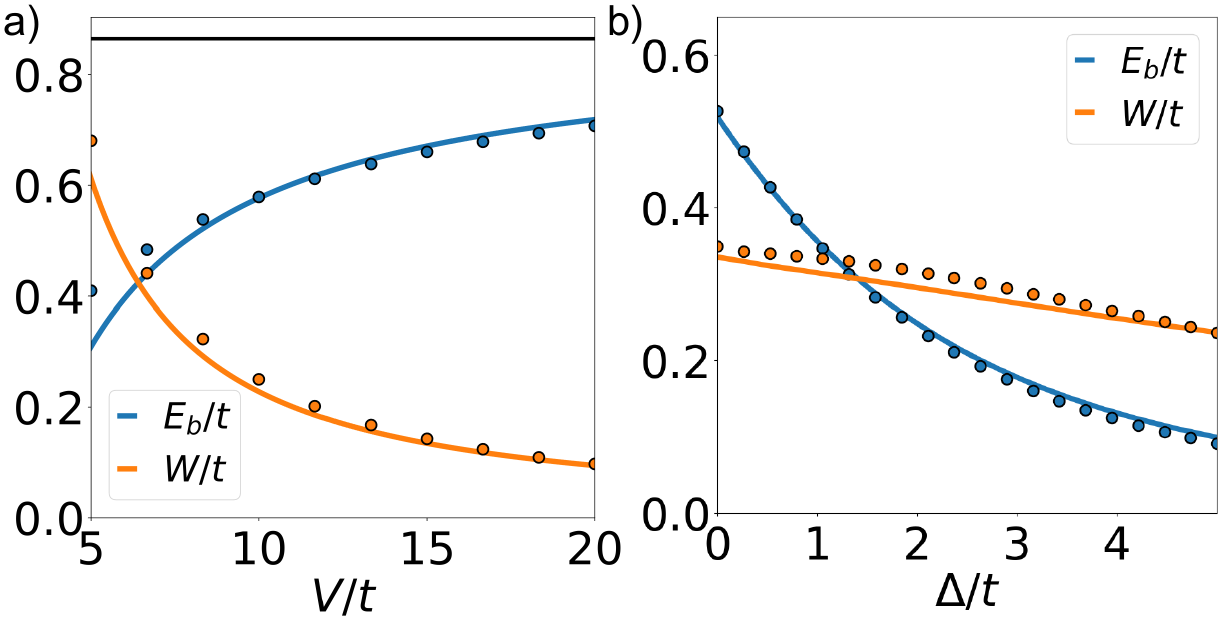}
        \vspace{-0.3cm}
    \caption{\textbf{Excitonic Cooper pair binding energy and bandwidth}: 
    Panel a) shows $E_b$ and $W$ as a function of $V/t$, respectively. The solid black line shows the asymptotic value {$\sqrt{3}t/2$} reached for $V/t=\infty$. Panel b) shows the evolution of $E_b$ and $W$ at $V/t=8$ as a function of $\Delta/t$. 24-site cluster ED (dots) and strong-coupling theory (solid lines) without any adjustable parameters.}
\label{fig:boson_Eb_W}
\end{figure}

Fig.~\ref{fig:boson_Eb_W} summarizes the main results on the behavior of excitonic Cooper pair in our system.  
Fig.~\ref{fig:boson_Eb_W}a) shows the evolution of the binding energy $E_b$ and the charge-$2e$ boson bandwidth $W$ as a function of the interaction strength $V$  for $\Delta=0$. As $V$ increases, the binding energy $E_b$ increases and eventually saturates, indicating the electronic origin of pairing from repulsion, while the boson bandwidth $W$ decreases because virtual processes entailed in boson hopping are energetically suppressed.   

Fig.~\ref{fig:boson_Eb_W}b) shows the effect of $\Delta$ on $E_b$ and $W$. 
The binding energy $E_b$ in Eq.~\eqref{2e_binding_energy} is reduced by increasing $\Delta$, which suppresses the hybridization of Cooper pair with exciton. On the other hand, at finite $\Delta$, the bosonic hopping amplitude takes the form: 
\begin{equation}\label{hopping_boson_delta}
    t_b = \frac{t^2}{6(\Delta +V)}\left[1+\frac{\Delta}{\sqrt{\Delta^2+12t^2}}\right]+\mathcal O\left(\frac{t^3}{V^2}\right),
\end{equation}
where higher order corrections are discussed in the SM~\cite{supplementary}. Thus, for sufficiently large $V/t$, the boson bandwidth first increases with $\Delta$ and then decreases when $\Delta$ becomes comparable to $V$. This opens up the possibility of using the displacement field as a tuning knob to crossover between different physical regimes. 

{\it Superconductivity and phase separation---} The formation of excitonic Cooper pairs from strong electron repulsion has important implications when our system is doped away from $\nu=1$, as we address below.  

The large binding energy $E_b \sim t$ gives rise to tightly bound pairs, which form charge-$2e$ bosons moving on a triangular lattice of $A$ sites with hopping $-t_b$. Depending on doping and microscopic details, bosons on a triangular lattice  exhibit different phases, including a superfluid phase where a boson condensate forms, phase separation, and a supersolid~\cite{Wessel_2005,Melko_2005,Auerbach_1997,Bernardet_2002}, which simultaneously develops a charge density wave and superfluidity.

In the following we present further evidence that these phases can be accessed within our model by tuning $\Delta$. Specifically, for small $\Delta$ and large $V/t$, we found in ED that excitonic Cooper pairs attract each other when placed on next-nearest-neighbor sites, as detailed in the SM~\cite{supplementary}. Therefore, at a small doping $\delta$, our system (in which interaction is short ranged \footnote{Phase separation is frustrated by the inclusion of longer-range Coulomb repulsion~\cite{kivelson_prl_2005}. Experimentally, the range of electron-electron interaction can be tuned by varying the distance of the sample to metallic gates.}) exhibits phase separation, where doped particles segregate into one phase at a high density $\nu'>1+\delta$ and the other phase is undoped $\nu=1$ insulator. 
{Indeed, our ED calculations of the ground state energy as a function of doping shows phase separation between $\nu=1$ and $\nu'=1\pm1/3$ in the infinite coupling limit $V\to\infty$ and $\Delta=0$, see SM~\cite{supplementary} for details.} 

On the other hand, upon increasing $\Delta$ at a fixed large $V/t$, we find that the interaction between excitonic Cooper pairs changes from attractive to repulsive above a critical value $\Delta^*$. For $V/t=8$, our ED calculation shows $\Delta^*/t\sim 1$. At $\Delta > \Delta^*$, our system with $\delta$ doped particles behaves as a two-dimensional Bose gas with repulsive interaction. Therefore, the ground state is a superfluid where charge-$2e$ bosons $b_r$ condense in the ${\bf q}=0$ state, leading to spin-polarized superconductivity from the condensation of excitonic Cooper pairs.  

At small doping, the superfluid density is small despite the large binding energy $E_b$. Thus, the superconducting critical temperature is governed by phase ordering~\cite{kosterlitz1973ordering,nelson_1977,Randeria1992,emery1995importance,Hazra2019}. To estimate $T_c$ of superconductivity of a gas of excitonic Cooper pairs, we employ the expression~\cite{PhysRevB.37.4936,PhysRevLett.87.270402,PhysRevLett.100.140405,zhang_Tc_2023,szhang_prx_2023,sous_prb_2024}: 
\begin{equation}\label{T_c}
    k_BT_c\approx C\frac{\hbar^2 \rho}{m_b} =C\frac{W}{3\sqrt{3}}|\delta|,
\end{equation}
where $\rho=|\delta|/(2\Omega)$ is the density of pairs with $\Omega=\sqrt{3}a^2/2$ unit cell area, and $W$ the bandwidth. 
In Eq.~\eqref{T_c}, $C$ depends very weakly on the repulsive interaction between bosons through a double log~\cite{PhysRevB.37.4936,PhysRevLett.87.270402,PhysRevLett.100.140405,supplementary}; we set $C\approx2\pi/\log(380/4\pi)$.

$T_c$~\eqref{T_c} depends linearly on the doping density $|\delta|$, in contrast to weak-coupling results where $T_c\propto\sqrt{\epsilon_F}$~\cite{Cr_pel_2021} and the Fermi energy $\epsilon_F$ is proportional to $|\delta|$. 
We note that in the absence of gate screening (i.e., for $1/r$ Coulomb interaction), the $T=0$ ground state of charged bosons at very low density is a Wigner crystal, whereas the superconducting state occurs above a critical density $r_s<60$ \cite{moroni_2004}. 
At temperatures above $T_c$ and below the binding energy $E_b$, we have a pseudogap regime where incoherent excitonic Cooper pairs constitute the charge carriers~\cite{Eagles1969,Leggett1980}.

The increase of $T_c$ with doping~\eqref{T_c} breaks down when the average distance between excitonic Cooper pairs shrinks to its size. This sets an upper bound on $T_c$, realized at boson density $\rho=1/(\pi\langle r^2\rangle)$ with $\langle r^2\rangle$ pair's mean square radius which corresponds to the filling factor $|\delta|=\sqrt{3}a^2/(\pi\langle r^2\rangle)\approx 0.55a^2/\langle r^2\rangle$.  
For $V/t$=8 and $\Delta/t>1$ (where bosons repel), our ED calculations show that $\langle r^2\rangle\approx 1.25a^2$, leading to a critical temperature $k_BT_c\approx0.06t$.
For a realistic hopping parameter $t=2.5${meV}, this results in $T_c=1.7$K.

{\it Discussion---} Among various mechanisms for superconductivity from repulsive interaction, the most widely studied is pairing due to spin fluctuation, especially near magnetic quantum critical points. Our work presents a diametrically opposite route to unconventional superconductivity. For fully spin-polarized systems, which are completely devoid of spin fluctuation, we show that electron pairing can arise upon doping from particle-hole fluctuations in a correlated insulator, and the underlying Cooper pair is strongly hybridized with the exciton.

Twisted $\Gamma$-valley TMDs are a promising platform for realizing our honeycomb lattice model. In this setting, strong Coulomb repulsion induces sublattice polarization~\cite{ma2024relativisticmotttransitionstrongly} and further drives ferromagnetism, as shown by our ED study of $\Gamma$-valley $t$TMDs and see also~\cite{Yang_2024,  zhang2024insulatingchargetransferferromagnetism,Devakul_2021}. 
This, in turn, establishes the parent state from which excitonic Cooper pair and  superconductivity may emerge at finite doping. A complementary approach to spin-polarized superconductivity in this system is discussed in Ref.~\cite{lina_future}.

Ferromagnetic superconductivity has been observed in rhombohedral graphene within both the spin-polarized, valley-unpolarized half-metal phase~\cite{Zhou_2021,Zhou_2022} and the valley-polarized quarter-metal phase~\cite{han2025signatureschiralsuperconductivityrhombohedral}. 
It will be interesting to explore the possibility of electron-hole fluctuations as a pairing mechanism, which can mediate inter-valley pairing in the half-metal state that corresponds to ${\bf q}=0$ and $f$-wave superconductivity in our model.

{\it Acknowledgments---} It is a pleasure to acknowledge useful discussions with Ahmed Abouelkomsan, Elio K\"onig, Pavel Volkov and Erez Berg. We are grateful to Kin Fai Mak and Jie Shan for private communications. We thank Lina Johnsen Kamra for related collaboration. 
This work was supported by Air Force Office of Scientific Research (AFOSR) under Award No. FA9550-22-1-0432. LF was supported in part by a Simons Investigator Award from the Simons Foundation.

\bibliography{biblio.bib}


\onecolumngrid
\newpage
\makeatletter 

\begin{center}
\textbf{\large Supplementary materials for:\\ ``\@title ''} \\[10pt]
Daniele Guerci$^{1}$ and Liang Fu$^{1}$ \\
\textit{$^1$Department of Physics, Massachusetts Institute of Technology, Cambridge, MA, USA}
\end{center}
\vspace{20pt}

\setcounter{figure}{0}
\setcounter{section}{0}
\setcounter{equation}{0}

\renewcommand{\thefigure}{S\@arabic\c@figure}
\makeatother

 \appendix

These supplementary materials contain the details of the continuum and tight-binding modeling of $\Gamma$-valley moir\'e semiconductors, exact diagonalization simulations and strong coupling perturbative results supporting our theory of exciton Cooper pairing in half metals.
Sec.~\ref{app:continuum_modeling_interacting_physics} provides a detailed discussion of the interacting properties of twisted $\Gamma$-valley semiconductors, including the continuum model, interaction effects, the resulting many-body physics and the tight-binding modeling.  
In Sec.~\ref{app:ed_lattice_model} we present details on exact diagonalization simulations of the tight-binding model. 
Sec.~\ref{app:infinite_coupling} focuses on the infinite coupling limit $V \to \infty$, presenting exact analytical results for one and two doped carriers, along with numerical studies of many-body physics involving $N$ doped carriers in the spin charge density wave ground state.
Sec.~\ref{app:strong_coupling} presents the analytical $1/V$ perturbation theory providing variational estimates to the boson dispersion relation, binding energy and ground state energy.  
Finally, analytical results valid in the ionic regime $\Delta\gg t$ are given in Sec.~\ref{app:field_theory}.

\section{Microscopic modeling of $\Gamma$-valley twisted semiconductors}
\label{app:continuum_modeling_interacting_physics}

$\Gamma$-valley moir\'e semiconductors are described by the continuum model introduced in Ref.~\cite{yangzhang_prb}:  
\begin{equation}\label{sm:hamiltonian_continuum}
    H(r)=-\frac{k^2}{2m}+\begin{pmatrix}
        u_t(r)+D/2 & t(r)  \\ 
        t(r) & u_b(r)-D/2
    \end{pmatrix},
\end{equation}
where in the previous expression $k=-i\hbar \nabla$, $u_{t/b}(r)=2V_0\sum_{j=1}^{3}\cos(g_j\cdot r\pm\phi)$, $t(r)=w_0+2w_1\sum_{j=1}^{3}\cos(g_j\cdot r)+2w_2\sum_{j=1}^{3}\cos(2g_j\cdot r)$ and $D$ is the displacement field. 
We fix the reference frame such that $a_j = a \exp[i\pi/2+2i\pi(j-1)/3]$ and $g_j=4\pi\omega^{j-1}/({\sqrt{3}a})$ with $\omega=\exp(2\pi i/3)$ and $a=a_0/(2\sin\theta/2)$ with $a_0$ the atomic lattice constant, where we utilized complex notation. 

We employed the parameters $w_0=338{\rm meV}$, $w_1=-16{\rm meV}$, $w_2=-2{\rm meV}$, $V_0=6{\rm meV}$ and $\phi=120^\circ$ derived for MoS$_2$ in Ref.~\cite{yangzhang_prb}, and the effective mass $m=0.8m_e$. Fig.~\ref{fig:bands_displacement_field} displays the bandstructure for different values of the displacement field $D$.

The large interlayer energy scale $w_0$ implies that the topmost bands are predominantly characterized by a layer-bonding configuration with only small layer imbalance $\gamma^z=\text{diag}[1,-1]$. As a result, the influence of the displacement field on the band structure is small when it is less than the interlayer bonding energy $w_0$, e.g. for a twist angle $\theta=2.876^\circ$ the sublattice gap is $1$meV for $D=30$meV and $3.5$meV for $D=100$meV. 

The model is invariant under the three-fold rotational symmetry $C_{3z}$, two-fold rotations $C_{2y}$, the three-dimensional inversion $\gamma^x H(-r)\gamma^x=H(r)$ with $\gamma^x$ Pauli matrix in the layer degree of freedom and $M_y$ mirror symmetry $y\to-y$. Moreover, the model preserves time-reversal symmetry ($\mathcal{T}$) and SU$(2)$ spin rotational symmetry, as the spin-orbit coupling is negligible at $\Gamma$~\cite{Korm_nyos_2015}. 
We note that $C_{2y}$ is broken either spontaneously (at filling $\nu=1$) when $D=0$ and above a critical interaction strength or explicitly when $D\neq0$ in the sublattice-polarized insulator. Additionally, in the spin-polarized regime—whether induced by interactions~\cite{Yang_2024,  zhang2024insulatingchargetransferferromagnetism} or an applied in-plane magnetic field—both $\mathcal T$ and SU(2)$_{\rm spin}$ symmetry are broken.

\subsection{Interacting continuum model and exact diagonalization results}

The many-body Hamiltonian reads: 
\begin{equation}
    H=\sum_{i} H(r_i) +\frac{1}{2}\sum_{i\neq j} V(r_i-r_j),
\end{equation}
where $H(r_i)$ is the single-particle Hamiltonian given in Eq.~\eqref{sm:hamiltonian_continuum}. We considered the double-gate screened Coulomb interaction, which in momentum space reads: 
\begin{equation}
    V(q)=\frac{e^2}{2\epsilon_0\epsilon}\frac{1}{q}\tanh d_{\rm sc}q,
\end{equation} 
where $d_{\rm sc}$ is the gate distance and $\epsilon$ the relative dielectric constant. The interacting physics is characterized by the competition of two energy scales: the kinetic energy $E_{\rm k} = \hbar^2/(2m a^2)$ and the interaction energy $E_{\rm int} = e^2 a_{}/(2\epsilon_0\epsilon  |a_1\times a_2|)$. Tendency to ferromagnetism is enhanced in the small twist angle regime, where the bandwidth is much smaller than the interaction energy scale. 
We perform exact diagonalization simulations projecting the Hamiltonian in the two topmost bands with dispersion $E_{kn}$:
\begin{equation}
    H=\sum_{k}\sum_n\sum_\sigma E_{kn} c^\dagger_{kn\sigma} c_{kn\sigma} + \frac{1}{2A}\sum_{\sigma_1\cdots\sigma_4}\sum_{n_1\cdots n_4}\sum_{k_1\cdots k_4}
    H_{q_1q_2,q_3q_4}
    c^\dagger_{k_1 n_1\sigma_1}c^\dagger_{k_2n_2\sigma_2}c_{k_3n_3\sigma_3}c_{k_4n_4\sigma_4},
\end{equation}
where $c_{kn\sigma}$ is the annihilation operator for an electron with momentum $k$, band index $n$ and spin $\sigma$, and $A=|L_1\times L_2|$. Furthermore, we have introduced the label $q_j=(k_j,n_j,\sigma_j)$ in the interaction matrix element. 
For a given set of indices the resulting matrix element reads  
\begin{equation}\label{matrix_element}
    H_{q_1q_2,q_3q_4}=\delta_{\sigma_1\sigma_4}\delta_{\sigma_2\sigma_3}\sum_g \delta_{k_1+k_2-k_3-k_4,\Delta g}\, V({k_1-k_4-g}) \,\Lambda^{n_1,n_4}_{k_1,k_4+g}\,\Lambda^{n_2,n_3}_{k_2,k_3+\Delta g-g},
\end{equation}
where we have introduced: 
\begin{equation}
    \Lambda^{n,m}_{k,p+g}=\int_{\rm UC}\frac{d^2r}{\Omega}e^{-ig\cdot r}u^*_{k n}(r) u_{pm}(r),
\end{equation}
where $\Omega=|a_1\times a_2|$, $\ket{u_{k\pm}}$ are the Bloch waves associated with the two low-energy bands hosting the Dirac cone.

\begin{figure}
    \centering
    \includegraphics[width=0.6\linewidth]{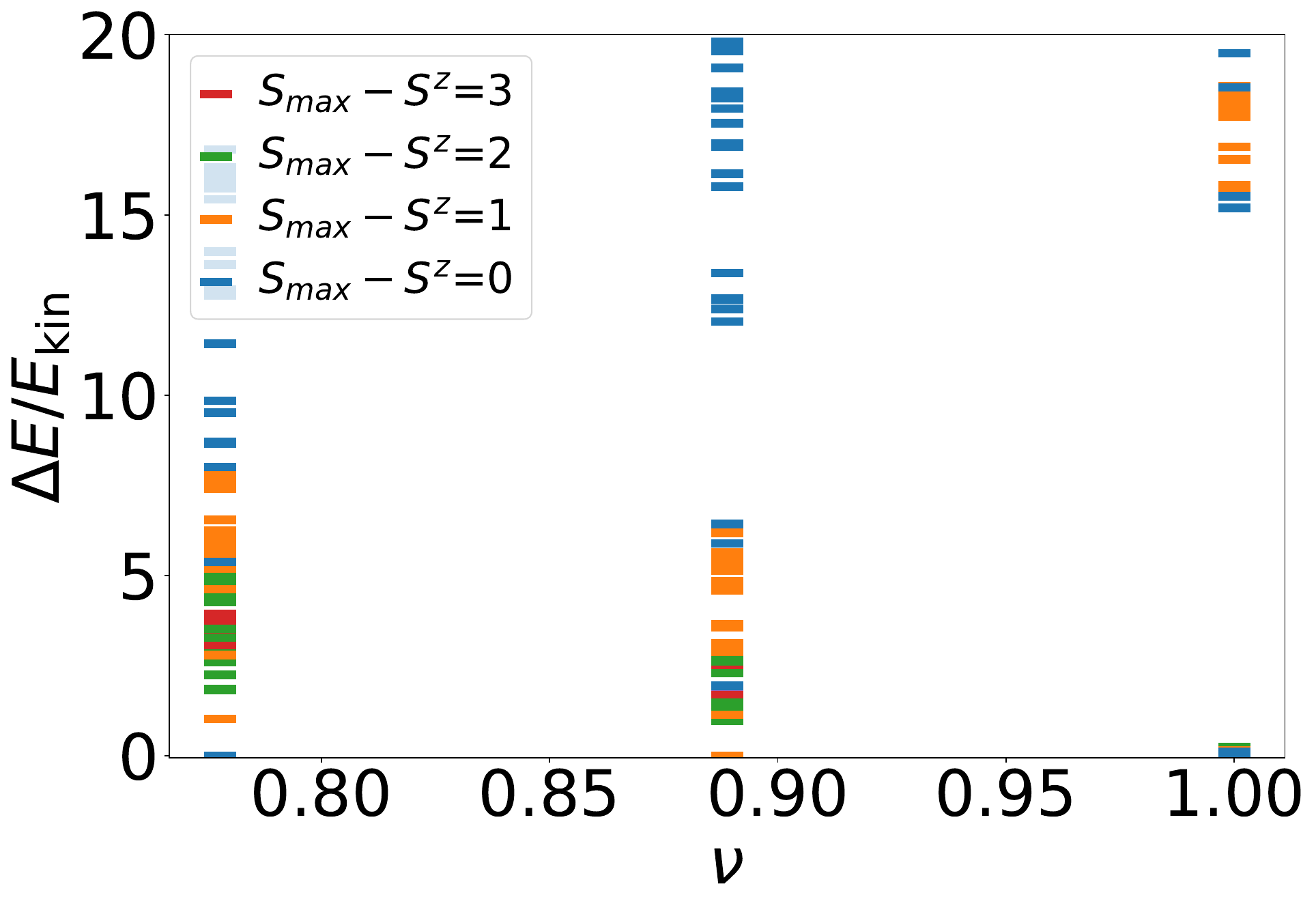}
    \caption{Many-body spectrum as a function of the filling factor $\nu$. The ground state sector features high-spin configurations evidence of tendency to form a ferromagnetic state. Calculations are performed setting $\theta=2^\circ$, $d_{\rm sc}=5$nm and $\epsilon=10$.}
    \label{fig:manybody_spectrum}
\end{figure}

\begin{figure}
    \centering
    \includegraphics[width=0.4\linewidth]{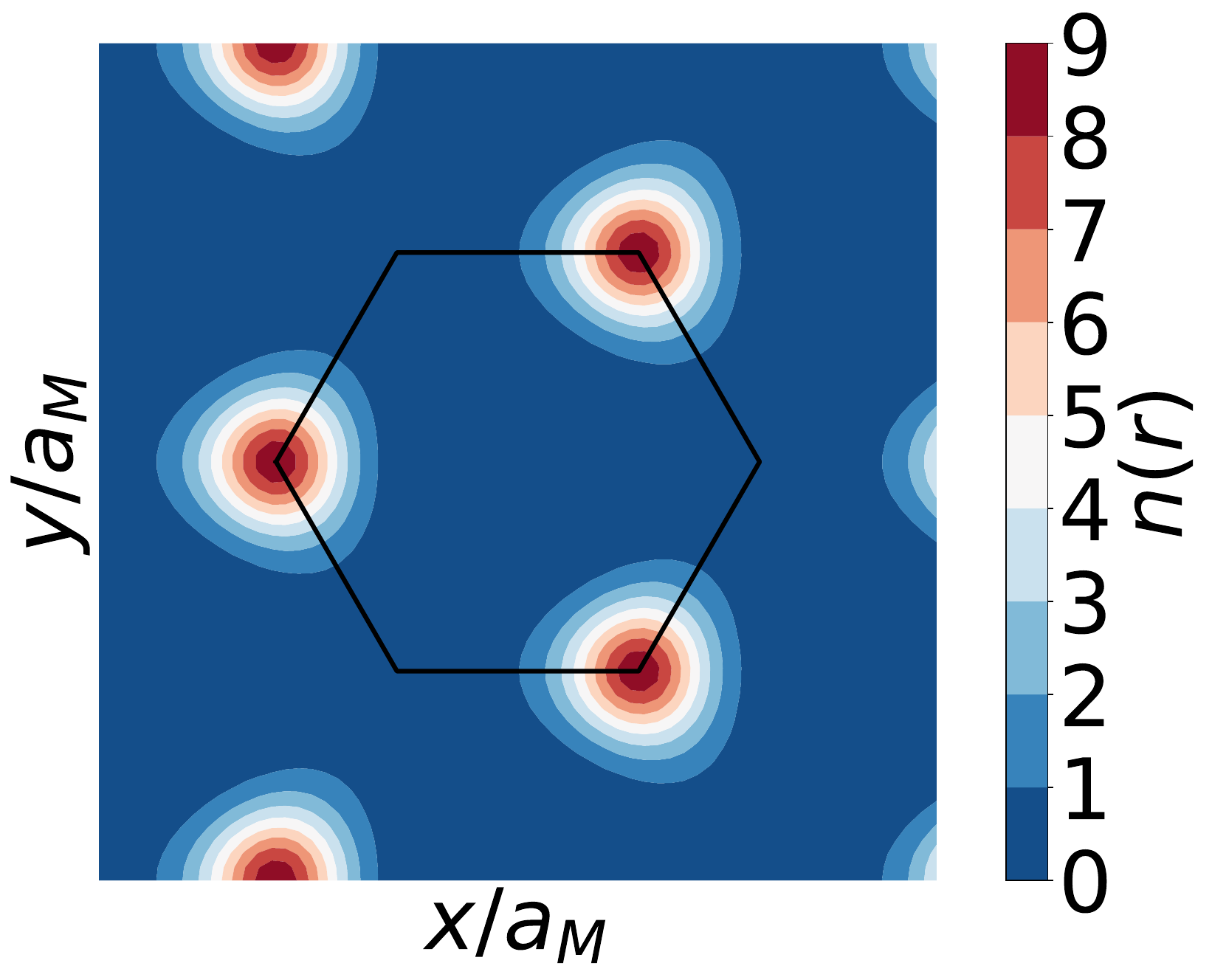}
    \caption{Charge distribution $n(r)=\mel{\Psi}{\psi^\dagger(r)\psi(r)}{\Psi}$ with $\ket{\Psi}$ the many-body ground state. Calculations are performed setting $\theta=2^\circ$, $d_{\rm sc}=5$nm, $\epsilon=10$ and a small displacement field $D$ ($D=.05$meV) applied to lift the degeneracy between the two ground states polarized on opposite sublattices. }
    \label{fig:charge_distribution_continuum}
\end{figure}

\begin{figure}
    \centering
    \includegraphics[width=.7\linewidth]{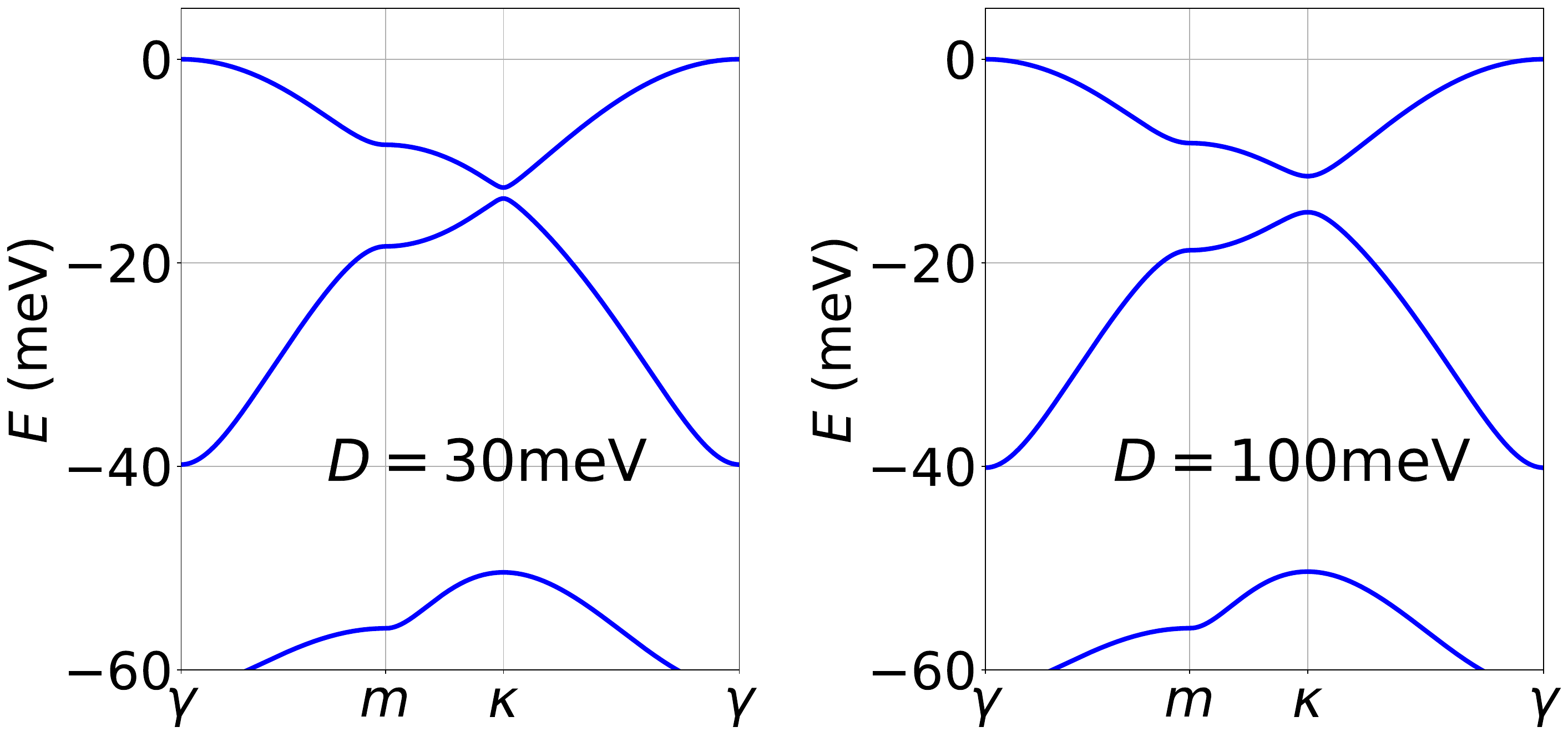}
    \caption{Bandstructure of gamma valley moir\'e semiconductors for $D=30,100$meV. Applying the displacement field breaks $C_{2y}$ and opens a trivial gap at $\kappa$ and $\kappa'$. The twist angle is given by $\theta=2.876^\circ$.}
\label{fig:bands_displacement_field}
\end{figure}

We perform exact diagonalization simulations on a $3 \times 3$ cluster that includes the two low-energy bands and both spin degrees of freedom. Fig.~\ref{fig:manybody_spectrum} shows the many-body spectrum for filling factors $\nu = 7/9$, $8/9$, and $1$, all exhibiting extensive ground-state degeneracy, reflected in the high-spin configurations that characterize the ground states. Interestingly, while the states at $\nu = 1$ and $7/9$ exhibit the full $(2S + 1)$ spin degeneracy, the $\nu = 8/9$ state—corresponding to a single doped hole—displays only a $(2S - 1)$ degeneracy. This reduction implies that the added hole forms a spin-singlet bound state, signaling the onset of polaronic physics. Furthermore, we observe that at filling factor $\nu=1$, the spin-polarized ground state sector is two-fold degenerate, with each ground state spontaneously breaking the $C_{2y}$ symmetry by localizing the charge distribution on one of the two sublattices as detailed in Fig.~\ref{fig:charge_distribution_continuum}.


\subsection{Limit of a strong moir\'e potential and sublattice basis}

\begin{figure}
    \centering
    \includegraphics[width=0.7\linewidth]{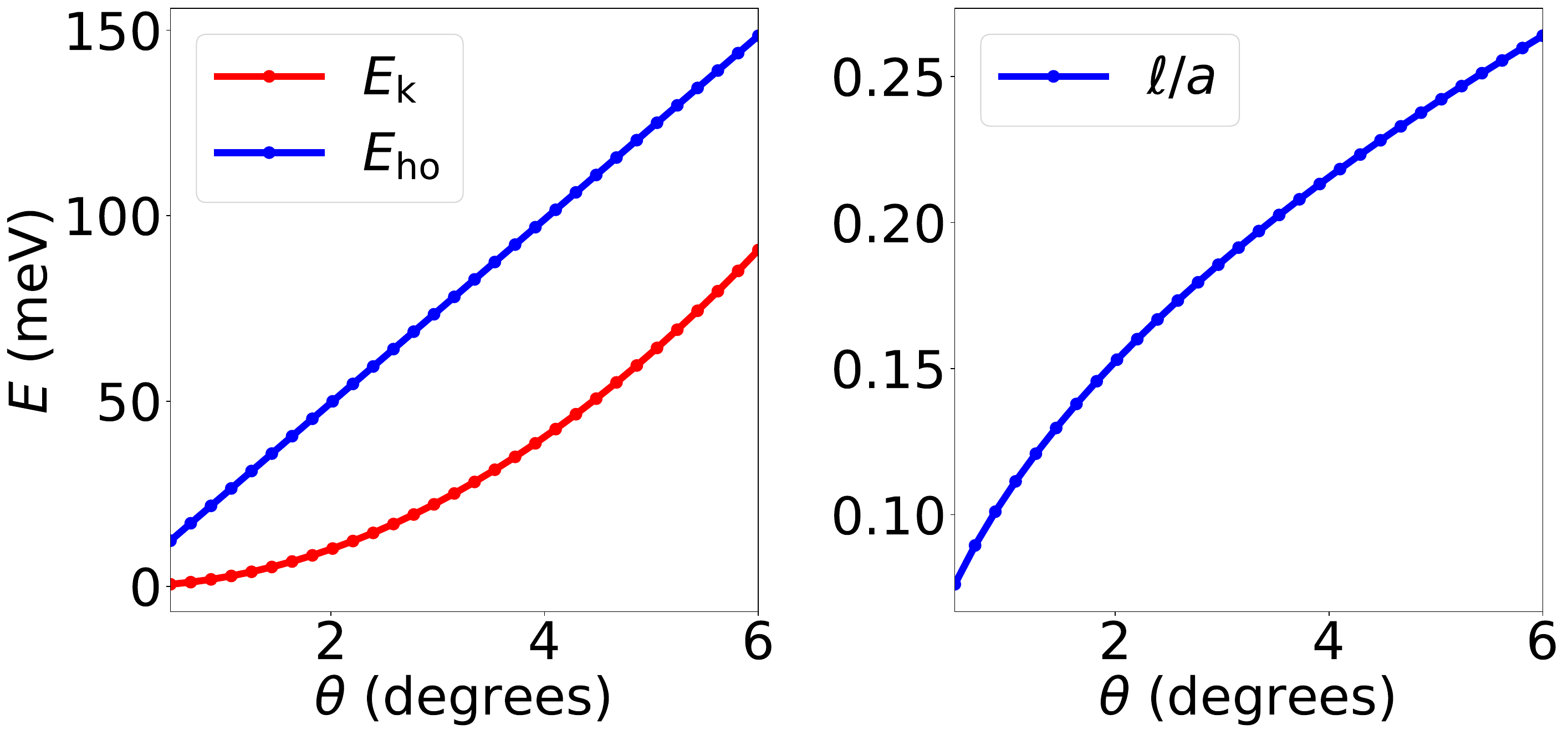}
    \caption{Energy scaling for quadratic dispersion (red) and harmonic oscillator (blue). Localisation length of the lowest energy state in each well. Notice that the distance between the two quantum well is $a/\sqrt{3}\approx 0.58 a$. }
    \label{fig:energy&lenght}
\end{figure}

In this section we take the limit of large moir\'e potential and we expand around the minima to determine the localization length of the orbitals. Our goal is to construct the basis that will be used to define the Wannier orbitals through projection~\cite{VanderbildtMarzari_prb,RevModPhys.84.1419}. The orbitals are localized around Wyckoff positions $\pm z_0=\pm(a_1-a_2)/3$ with $a_{1/2}=e^{-i\pi/6},e^{i\pi/2}$. Expanding around these points we find: 
\begin{equation}u_b(z_0+\delta r)\approx 6V_0 -\frac{4\pi^2 (2V_0)}{a^2}\delta r^2,\quad u_t(z_0+\delta r)\approx -3V_0+\frac{4\pi^2 V_0}{a^2}\delta r^2,
\end{equation}
in addition we have the expansion of the potential $\Delta(r)$: 
\begin{equation}
    \Delta(z_0+\delta r)\approx \bar w +\frac{4\pi^2}{a^2}\left(w_1+4w_2\right)\delta r^2,
\end{equation}
$\bar w=w_0-3(w_1+w_2)=392$meV and $\Delta w=w_1+4w_2=-24$meV. We observe that the difference in $\epsilon_{b/t}$ introduces an asymmetry between $\pm z_0$ high symmetry stackings leading to a finite out-of-plane polarization of the orbitals resulting in a net response to an applied displacement field. Around $z_0$, the Hamiltonian is approximated as: 
\begin{equation}\label{moire_potential}
    H(r+z_0)\approx \frac{9V_0}{2}\sigma^z + \bar w \sigma^x -\frac{k^2}{2m}+\frac{4\pi^2 \delta r^2}{a^2}\begin{pmatrix}
        -2V_0 & \Delta w \\ 
        \Delta w & V_0
    \end{pmatrix},
\end{equation}
where $9V_0/(2\bar w)\approx 0.07\ll 1$ implying that the state is mostly described by a layer symmetric configuration. The eigenstates diagonalizing the potential are given by: 
\begin{equation}\label{sm:layer_distribution}
    \ket{v_+} = [\cos\chi/2,\sin\chi/2]^T,\quad \ket{v_-}=[-\sin\chi/2,\cos\chi/2]^T,
\end{equation}
with $\chi=\arctan\Delta_x/\Delta_z$ and $\Delta_x=\bar w $, $\Delta_z=9V_0/2+D/2$ if a displacement field is applied. 
The energy gap between the two states $v_{\pm}$ is large and we perform projection to the topmost state $\ket{v_+}$. Notice that $\mel{v_+}{\bm\tau}{v_+}=(0.9976,0,0.0687)$ is remarkably close to the layer distribution obtained from the Bloch state at $z_0$: $\mel{\bm \psi_{\gamma 1/2}(z_0)}{\bm \tau}{\bm \psi_{\gamma 1/2}(z_0)}=(0.9980,0,0.0624),(0.9980,0,0.0635)$. We proceed simply projecting the space dependent part of the Hamiltonian in the topmost configuration obtaining: 
\begin{equation}
    \mathcal H_+\equiv \mel{v_+}{H(r+z_0)}{v_+} = -\frac{k^2}{2m} - \frac{4\pi^2\delta r^2}{a^2}E_h,\quad  E_h=\mel{v_+}{\begin{pmatrix}
        -2V_0 & \Delta w \\ 
        \Delta w & V_0
    \end{pmatrix}}{v_+}.
\end{equation}
The Hamiltonian in the hole like picture can be then written as: 
\begin{equation}
    \mathcal H_+ = \frac{\hbar^2k^2}{2m}+\frac{\alpha r^2}{2a^2} 
\end{equation}
with $\alpha=8\pi^2E_h$. We readily find the frequency of the harmonic oscillator $\omega$, the localisation length $\ell$ and the wavefunction $\psi_0$:
 \begin{equation}
\hbar \omega =\sqrt{\frac{8\pi^2\hbar^2}{ma^2}E_h},\quad \frac{\ell}{a} = \left[\frac{\hbar^2 }{8\pi^2 ma^2E_h}\right]^{\frac{1}{4}},\quad  \psi_0( r) =\frac{e^{-r^2/(2\ell^2)}}{\ell \sqrt{\pi}}.
 \end{equation}
The spread of the wavefunction decays with the square root of the twist angle as the twist angle is reduced as shown in Fig.~\ref{fig:energy&lenght}. Therefore, we have two different low-energy localised states given by: 
\begin{equation}\label{gaussian_basis}
    \bm\psi_+( r) =\ket{v_{+,+}}\frac{e^{-( r - z_0)^2/(2\ell^2)}}{\ell\sqrt{\pi}},\quad \bm\psi_-(r) =\ket{v_{+,-}}\frac{e^{-(r-z_1)^2/(2\ell^2)}}{\ell\sqrt{\pi}},
\end{equation}
where $z_1=-\omega z_0$ and $\ket{v_{+,\pm}}$ the topmost energy eigenstates diagonalizing the moir\'e potential around $z_{0/1}$~\eqref{moire_potential}.
\subsection{Wannier orbitals via projection}

We now employ the states $\bm\psi_{\pm}$~\eqref{gaussian_basis} to build the Wannier functions obtained of the two topmost bands. To this aim we start from the Bloch orbitals consisting of two component spinors $\bm \psi_{k}=[\psi_{k t},\psi_{k b}]^T$ in the layer index $\ell=t,b$: 
\begin{equation}
    \bm\psi_{k n}(r) =e^{ik\cdot r}  \sum_{g} \bm z_{n g}(k) e^{ig\cdot r},
\end{equation}
where $g=n g_1+mg_2$ with $n,m\in\mathbb Z$ are reciprocal lattice vectors and $\bm z_{n g}(k)$ a spinor in the layer degree of freedom.  We employ the exact expressions for the wavefunctions in the large potential limit projecting the Bloch states into the set of states $\bm\psi_\pm( r)=\ket{v_{+,\pm}}f_{\pm}( r)$ where $f$ is a Gaussian function centered at the potential minima~\eqref{gaussian_basis}.  
To this aim we first introduce the projected basis: 
\begin{equation}\begin{split}
   \bm \phi_{k n}( r)& = \sum_{m} \bm \psi_{k m}( r)  \braket{\psi_{k m}}{\psi_n}=\sum_m \bm \psi_{k m}( r) \sum_{ g} \braket{z_{m g}(k)}{v_{+,n}}\int d^2 r e^{-i(k+ g)\cdot r}f_{n}( r)\\
   & = \sum_m \bm \psi_{k m}( r) \sum_{ g} \braket{z_{m g}(k)}{v_{+,n}} e^{-i(k+ g)\cdot z_n} e^{-(k+ g)^2\ell^2_n/2}.
\end{split}\end{equation}
The new eigenstate basis $\bm\phi_{kn}$ serves as the starting point for constructing the Wannier function:
\begin{equation}\label{auxiliary_orbitals}
    \tilde{\bm\psi}_{k n}(\bm r) = \sum_{j} \bm \phi_{k j}( r)S^{-1/2}_{jn}(k), \quad S(k) = A^\dagger(k) A(k),
\end{equation}
and we have introduced the matrix: 
\begin{equation}
    A_{mn}(k)=\sum_{\bm g} \braket{z_{m\bm g}(k)}{v_{+,n}} e^{-i(k+\bm g)\cdot z_n} e^{-(k+\bm g)^2\ell^2_n/2}.
\end{equation}
Notice that Eq.~\eqref{auxiliary_orbitals} can be also expressed as: 
\begin{equation}\label{sm:unitary_transformation}
    \tilde{\bm\psi}_{k n}(\bm r) = \sum_{j} \bm \psi_{k j}( r)U_{jn}(k), \quad U(k) = A(k) S^{-1/2}(k) 
\end{equation}
with $U(k)$ unitary matrix. 
Finally, the Wannier orbitals are given by: 
\begin{equation}
   \bm  W_{ R n}( r) =  \frac{1}{N}\sum_{k}\tilde{\bm\psi}_{k n}( r) e^{-ik\cdot R}=\frac{1}{N}\sum_{k}\tilde{\bm\psi}_{k n}( r) e^{-ik\cdot R},
\end{equation}
with $N$ number of unit cells and $R=n a_1+ma_2$ with $n,m\in\mathbb Z$. 

\begin{figure}
    \centering
    \includegraphics[width=.7\linewidth]{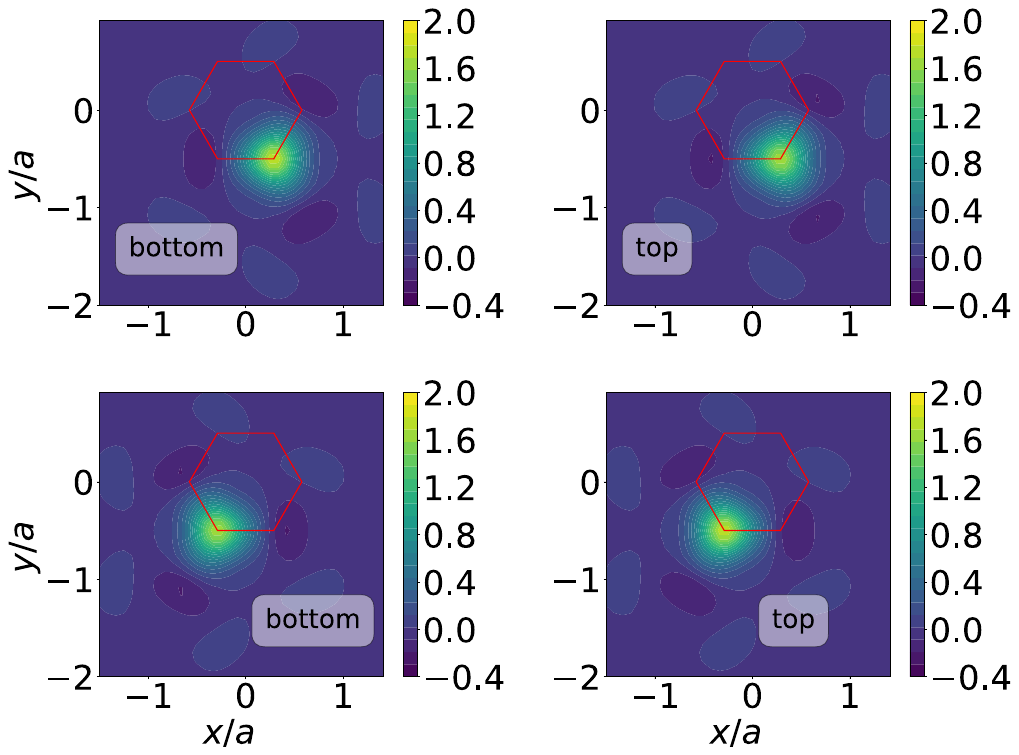}
    \caption{Layer resolved Wannier functions for the two topmost bands of gamma valley twisted semiconductors for twist angle $\theta=2.876^\circ$ forming a honeycomb lattice. The left and right columns represent the bottom and top layers, respectively, while the upper and lower rows depict the Wannier functions $\bm W_{RA/B}(r)$ localized at the MX and XM high-symmetry stacking configurations, which form the two sublattices $A/B$ of the emergent honeycomb lattice.}
    \label{fig:sublattice_basis}
\end{figure}

The Wannier orbitals are shown in Fig.~\ref{fig:sublattice_basis} for the twist angle $2.876^\circ$ and $D=0$. The Wannier orbitals show a slight layer imbalance ($\mel{\bm W_{RB}}{\gamma^z}{\bm W_{RB}}=-\mel{\bm W_{RA}}{\gamma^z}{\bm W_{RA}}\neq0$) with $\gamma^z=\text{diag}[1,-1]$ in the layer degree of freedom and transform into each other under $C_{2y}$.

\subsection{Tunneling amplitudes, minimal lattice model and Coulomb repulsion} 

\begin{figure}
    \centering
    \includegraphics[width=.9\linewidth]{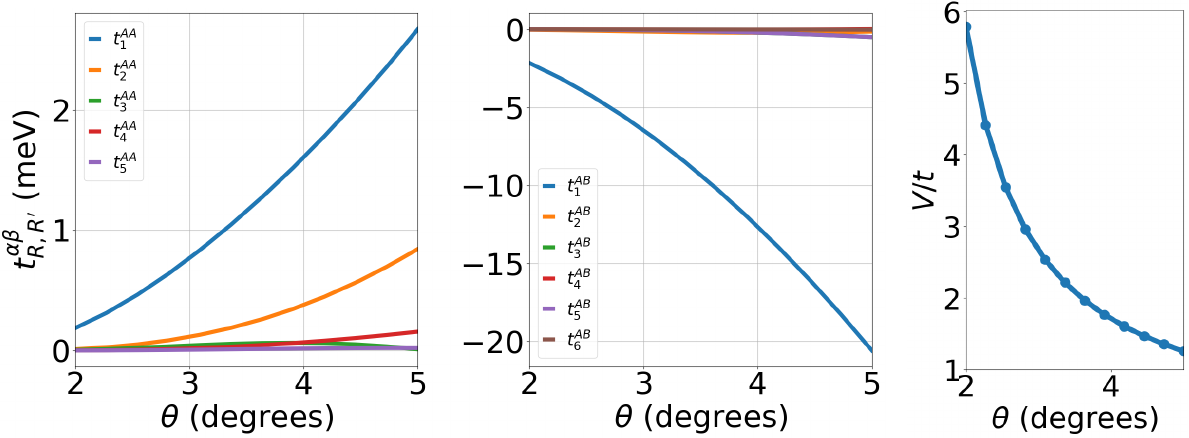}
    \caption{Left panels show Tunneling amplitudes for vanishing displacement field $D=0$. The right most panel shows the ratio $V/t$ with $V$ and $t$ nearest-neighbor interaction and hopping, respectively. In this case, $\Delta=0$ and $t^{AA}_n=t^{BB}_n$. }
    \label{fig:tunneling_angle}
\end{figure}
\begin{figure}
    \centering
    \includegraphics[width=1\linewidth]{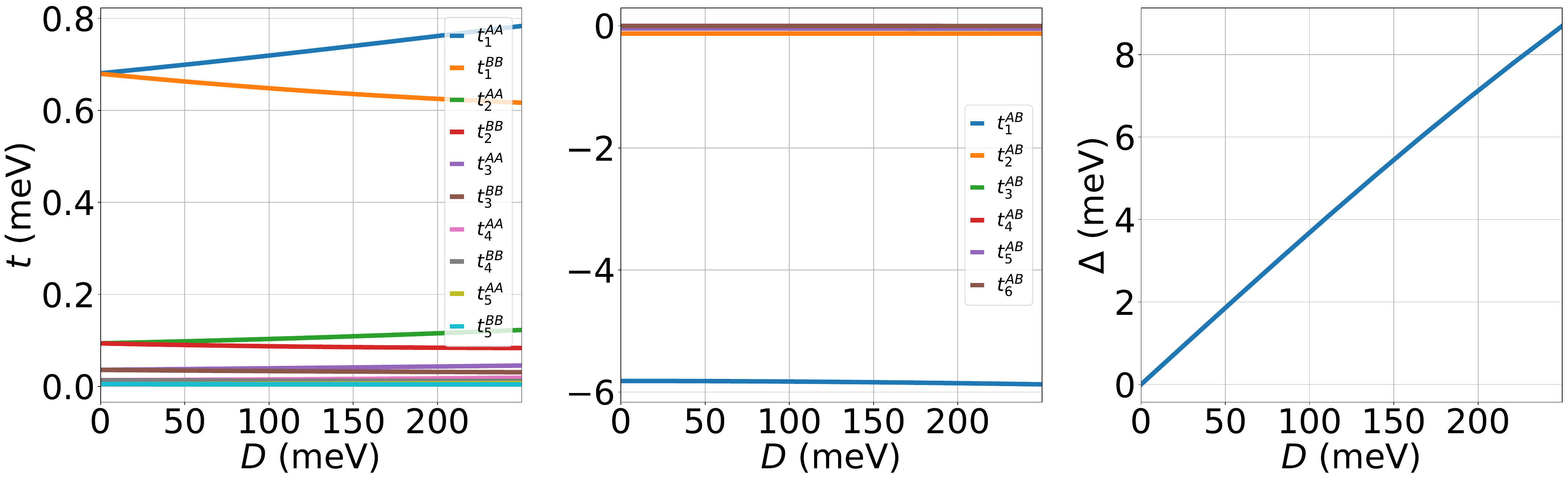}
    \caption{Evolution of tight binding parameters with the displacement field $D=0$ for $\theta=2.876^\circ$. The leading hopping process is the n.n. tunneling connecting the two sublattice. The right panel shows the evolution of $\Delta$ as a function of $D$. Due to the large interlayer energy scale $w_0$ given below Eq.~\eqref{sm:hamiltonian_continuum}, the wavefunctions exhibit only a slight interlayer imbalance.}
    \label{fig:tunneling_delta}
\end{figure}
We now compute the hopping amplitudes: 
\begin{equation}
    t^{n,n'}_{R,R'} = \frac{1}{N} \sum_{k}\sum_{l=0,1}e^{ik\cdot(R-R')} U^\dagger_{nl}(k)E_{k l}U_{ln'}(k),
\end{equation}
where $l$ extends only to the topmost twofold manifold of bands and $U(k)$ is the unitary transformation to the sublattice basis given in Eq.~\eqref{sm:unitary_transformation}. 

The evolution of the leading hopping terms $t^{AB}_n\equiv t^{AB}_{a_n,0}$, $t^{BB}_n=t^{BB}_{a_n,0}$,  and $t^{AA}_n=t^{AA}_{a_n,0}$ as a function of the twist angle is given in Fig.~\ref{fig:tunneling_angle} with subscript denoting the increasing number of shell in real space. For the intrasublattice hopping $t^{AA},t^{BB}$ the contribution of the higher shell decreases with increasing distance on the lattice, while the intersublattice hopping is dominated by the n.n. contribution.
Furthermore, Fig.~\ref{fig:tunneling_delta} shows the evolution of the hopping and sublattice gap as a function of the applied displacement field. The hopping $t^A_2$ and $t^B_2$ change differently for the two different sublattice degrees of freedom by increasing the displacement field $D$. Specifically, the state with smaller zero point energy is less confined, has a larger localization length and, therefore, a larger hopping amplitude. 
In momentum space the lattice model describing twisted $\Gamma$ twisted homobilayers is in the basis $\Psi_k=[a_k,b_k]$: 
\begin{equation}\label{app:tight-binding}
    H_k = \begin{pmatrix}
        \epsilon_{kA} +\Delta/2 & t_k\\
        c.c. & \epsilon_{kB} -\Delta/2,
    \end{pmatrix}
\end{equation}
with $a_j$ lattice vectors and $u_j$ connecting the two different sublattices. We emphasize again that a finite displacement field introduces a finite gap $\Delta$ and also modifies the hopping $t^{AA/BB}_n$ for the two different sublattices. A general expression for the intrasublattice dispersion and the interlayer tunneling read: 
\begin{equation}
    \epsilon_{k a} = 2\sum_{n} t^a_n \sum_{j\in R_n}\cos k\cdot(x_{nj}a_1+y_{nj}a_2 ), \quad t_{k}=\sum_n t^{AB}_n\sum_{j\in R_n}e^{ik\cdot(u_1+x_{nj}a_1+y_{nj}a_2)},
\end{equation}
with $R_n$ identifying the $n$-th shell of lattice sites. 

The leading hopping amplitude is the intersublattice n.n. amplitude $t^{AB}_1$, which including the sublattice potential $\Delta$ and the n.n. repulsion $V=e^2/(4\pi\epsilon_0\epsilon a)$ define the minimal model investigated via exact diagonalization. The ratio between $V$ and $t$ is  shown in the right panel of Fig.~\ref{fig:tunneling_angle}.

\subsection{Strong coupling Hamiltonian: spin model at filling 1}

\begin{figure}
    \centering
    \includegraphics[width=.9\linewidth]{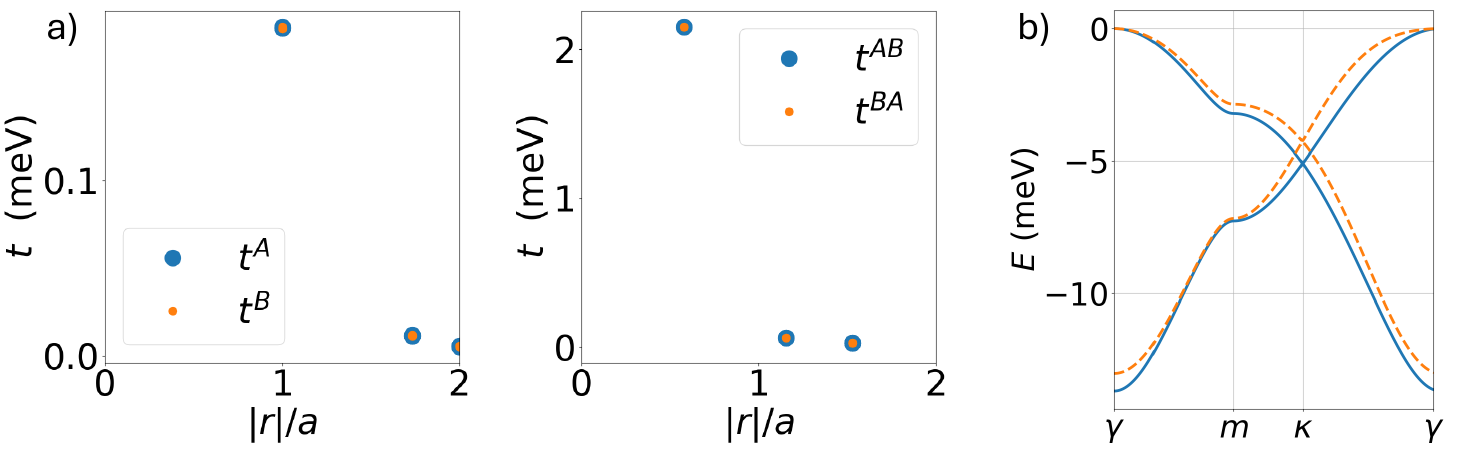}
    \caption{Panel a) shows the intralayer $t^{A/B}$ (left) and interlayer $t^{AB/BA}$ (right) hopping amplitudes as a function of the distance between sites $|r|$. Panel b) shows a comparison between the continuum model and tight-binding bands obtained including only hopping between sites up to $|r|\leq a$. Results are obtained setting $\theta=2^\circ$. }
    \label{fig:tight_binding_continuum}
\end{figure}

Additional insights into the magnetic properties can be gained through the tight-binding model~\eqref{app:tight-binding} derived from the continuum model. In the small twist-angle regime, the interacting tight-binding model reads:
\begin{equation}\label{app:t_tp_hopping}
    H = t\sum_{\langle r,r'\rangle}c^\dagger_{r} c_{r'} +t'\sum_{\alpha=A,B}\sum_{\langle r,r'\rangle\in\alpha} c^\dagger_r c_{r'} + U\sum_{r}n_{r\uparrow}n_{r\downarrow}+V\sum_{\langle r,r'\rangle}n_r n_{r'},  
\end{equation}
where $t$ is the nearest-neighbor and $t'$ the next-nearest neighbor hoppings with $t'/t=0.09$ and $t=2$meV for $\theta=2^\circ$. The hopping parameters and a comparison between the tight-binding Hamiltonian including $t$ and $t'$~\eqref{app:t_tp_hopping}, and continuum model band structures are shown in Fig.~\ref{fig:tight_binding_continuum}.

In the sublattice-polarized insulator at $\nu = 1$, where the charge degrees of freedom are localized on a triangular lattice—either sublattice $A$ or $B$—the effective spin-exchange Hamiltonian takes the form:
\begin{equation}
    H_{\rm spin} = J\sum_{\langle r,r'\rangle\in \alpha}\bf S_{r}\cdot \bf S_{r'}.
\end{equation} 
Here, $J = 4t'^2/U - t't^2/(V + \Delta/2)^2$, where the first (antiferromagnetic) contribution arises from the leading exchange processes, including superexchange, while the second (ferromagnetic $t'>0$ in Fig.~\ref{fig:tight_binding_continuum}) term originates from loop exchange processes~\cite{Devakul_2021,zhang2024insulatingchargetransferferromagnetism}.
In the small twist angle regime and for screened Coulomb interactions, where $U$ constitutes the dominant energy scale and $t'/t\le0.1$, leading to $J < 0$.

\section{Exact Diagonalization Calculations of the lattice model}
\label{app:ed_lattice_model}

We perform exact diagonalization calculations of the hexagonal lattice model with parameters $t,V$ and $\Delta$: 
\begin{equation}\label{hamiltonian_minimal}
    H = -t\sum_k \left(f_k \psi^\dagger_{kA} \psi_{kB} + h.c. \right)-\Delta \frac{\psi^\dagger_{kA} \psi_{kA}-\psi^\dagger_{kB} \psi_{kB}}{2}+\frac{V}{N}\sum_{\{k\}}\delta_{k_1+k_2,k_3+k_4}f(k_3-k_2)\psi^\dagger_{k_1A}\psi^\dagger_{k_2B}\psi_{k_3B}\psi_{k_4A},
\end{equation}
the form factor is $f_k = \sum_{j=1,2,3} e^{ik\cdot u_j}$ and $u_j= e^{2i\pi(j-1)/3}/\sqrt{3}$ in complex notation. 
These parameters can be derived explicitly from the knowledge of the Wannier orbitals. Additionally, we have introduced the fermionic operators: 
\begin{equation}
    \psi_{kA}=\frac{1}{\sqrt{N}}\sum_{r\in A} e^{ik\cdot r}f_r,\quad \psi_{kB} = \frac{1}{\sqrt{N}}\sum_{r\in A} e^{ik\cdot (r+u_1)}f_{r+u_1}.
\end{equation}
In our numerical simulations, we consider a $3\times4$ cluster ($24$ sites in total) with periodic boundary conditions.

\begin{figure}
    \centering
    \includegraphics[width=0.6\linewidth]{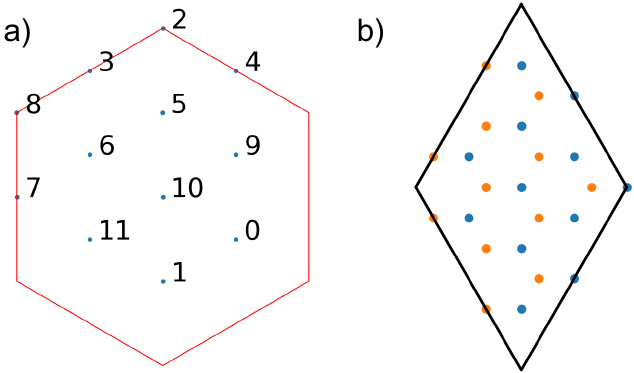}
    \caption{Momentum grids in the BZ for 12 corresponding to $24$ sites in total. }
    \label{fig:momentum grid}
\end{figure}
 
\subsection{Incompressible state at filling $\nu=1$}

\begin{figure}
    \centering
    \includegraphics[width=.8\linewidth]{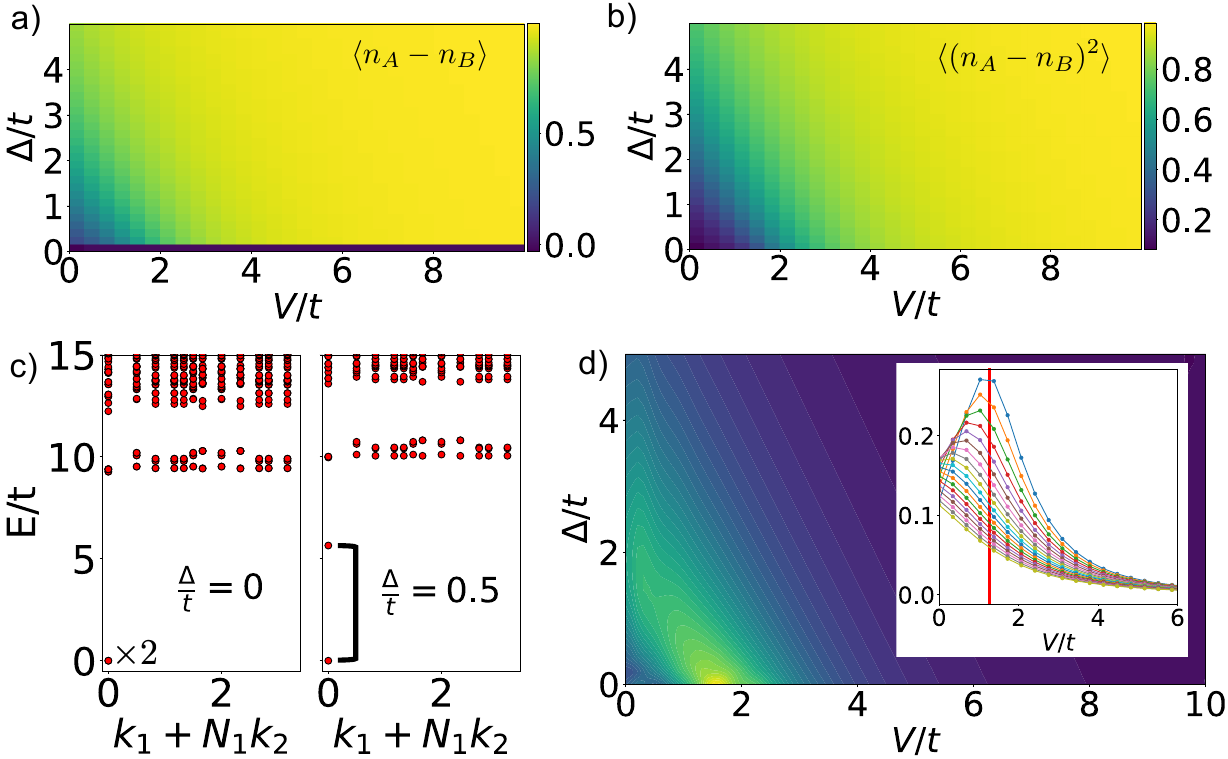}
    \caption{Panel a) and b) show $\langle n_A-n_B\rangle$ and $\langle(n_A-n_B)^2\rangle$ in a relevant range of $\Delta/t$ and $V/t$. Panel c) illustrates the many-body spectrum for $V/t=5$ and two different values of $\Delta/t=0$ (left) and $\Delta/t=0.5$ (right). Panel d) shows the derivative of $\langle(n_A-n_B)^2\rangle$ with respect to $\Delta$. }
    \label{fig:filling_1}
\end{figure}
In this section, we detail the properties of the parent state of the superconductor. This state is a sublattice-polarized insulator, where the symmetry between the two sublattices is broken. The symmetry breaking occurs spontaneously at $\Delta=0$ and explicitly for any finite $\Delta$. Fig.~\ref{fig:filling_1}a) and~\ref{fig:filling_1}b) show the expectation values $\langle n_A-n_B\rangle$ and $\langle(n_A-n_B)^2\rangle$ with $n_{A/B}=N_{A/B}/N$. The increase of $\langle(n_A-n_B)^2\rangle$ signals the system's approach toward a strongly correlated sublattice insulator. Fig.~\ref{fig:filling_1}c) displays the many-body spectrum at filling $\nu=1$ for two different values of $\Delta/t$. The two-fold degeneracy of the ground state at $\Delta=0$ signals the degeneracy of the two sublattice insulating states with opposite sublattice polarization that are connected by $C_{2y}$. Introducing $\Delta$ lifts the degeneracy opening a gap of approximately $N_s\Delta$ growing extensively with system size. 
Finally, Fig.~\ref{fig:filling_1}d) shows the derivative of $(n_A-n_B)^2$ with respect to $\Delta$. The inset of Fig.~\ref{fig:filling_1}d) shows a peak at $V_c/t=1.3$ (solid red line) that we interpret as the location of the Ising Gross-Neveu critical point~\cite{gross_1974} at zero displacement field.

\subsection{Computing properties of the $2e$ bound state: effective mass, mean square radius and binding energy}

In this section we detail the evaluation of the binding energy, effective mass and the mean square radius. To simplify the notation we define $E_p$ the ground state energy in the sector with $N+p$ particles. 
The binding energy per particle is defined as: 
\begin{equation}\label{binding_energy}
    E_b = E_{1e}-E_{2e}/2,
\end{equation}
where: 
\begin{equation}
    E_{2e} = E_{2}- E_{0},\quad E_{1e} = E_{1}- E_{0}.
\end{equation}
$E_{2}$ lies in the sector with $Q=0$ (mod reciprocal lattice vectors), while $E_{1}$ is defined as: 
\begin{equation}
    E_{1} \equiv \min_{Q} E_{1}(Q),
\end{equation}
where the minimum is found scanning over different momentum sectors. 
In agreement with perturbative results, we find that the minimum in the particle sector $1$ is always located at $K/K'$. Finally, the charge gap is defined as: 
\begin{equation}
    E_{\rm gap} = E_1+E_{-1}-2E_0.
\end{equation}

\begin{figure}
    \centering
    \includegraphics[width=0.75\linewidth]{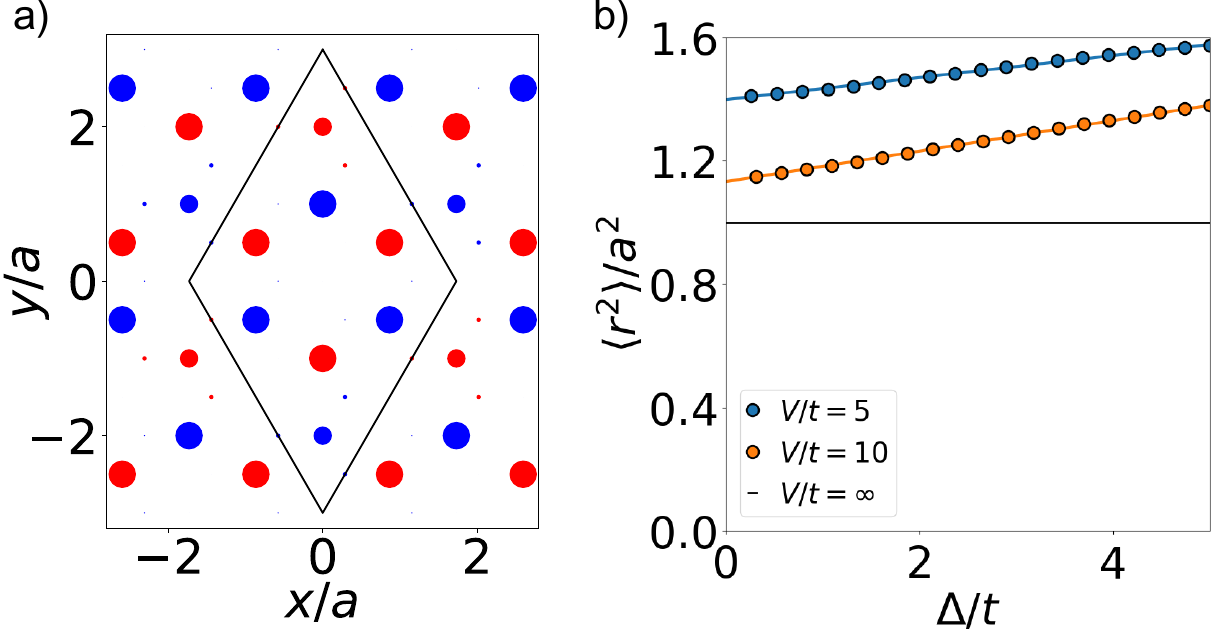}
    \caption{Panel a) $2e$ bound state wavefunction on a torus $(L_1,L_2)$ including $12$ unit cells for $V/t=10$ and $\Delta/t=0.8$. The size of the dots show $|\Psi_{2e}|$ and the color represents the phase blue ($0$) and red($\pi$). Panel b) presents the mean square radius, with dotted lines showing the power-law fit $\langle r^2\rangle/a^2=A\Delta/t+B$ obtained in the long wavelength limit. }
    \label{fig:2e_wavefunction}
\end{figure}

We compute the mass of the pair looking at the dispersion in the many-body space with $2$ particles: 
\begin{equation}\label{inverse_mass}
    \frac{1}{m_B} = \frac{\nabla^2_Q E_{2}(Q)}{2}\Bigg|_{Q=0},
\end{equation}
where the isotropy of the mass follows from the $C_{3z}$ symmetry of the theory. Given the $D_6$ symmetric cluster in Fig.~\ref{fig:momentum grid} the second derivative~\eqref{inverse_mass} can be approximated employing the first shell of momenta: 
\begin{equation}
    \frac{1}{m_b}=\frac{1}{2}\sum_{l}\frac{E_{N+2}(Q=\Delta k_l)-E_{N+2}(\Gamma)}{\sum_{j=1}^3 (1-\cos\Delta k_l\cdot a_j)}.
\end{equation}
Finally, the mean square radius $\langle r^2\rangle$ is obtained by first computing the two particle bound state wavefunction 
\begin{equation}\label{2ewavefunction}
    \Psi_{2e}(x+\Delta x ,x)=\mel{\Psi_N}{f_{x+\Delta x }f_x}{\Psi_{N+2}},
\end{equation}
over the ground state obtained via exact diagonalization. The latter average value reads: 
\begin{equation}
     \mel{\Psi_{N}}{f_{x+\Delta x }f_x}{\Psi_{N+2}} = \frac{1}{N}\sum_{kk'} e^{ik\cdot(x+\Delta x)+i k'\cdot x}\mel{\Psi_0}{\psi_{k\alpha} \psi_{k'B}}{\Psi_{2e}}, 
\end{equation}
where $\alpha=A,B$ depending on the sublattice site and momentum conservation selects the amplitudes with $k+k'=0$ mod reciprocal lattice vectors $\mod\left[l_j\cdot (k+k'),2\pi\right]=0$. Fig.~\ref{fig:2e_wavefunction}a) shows the $2e$ wavefunction where the size of the dots represents the absolute value of the wavefunction and the colorcode the phase. Finally, we observe that the wavefunction is finite but small in the $A$-sublattice, with average occupation number $\langle n_A\rangle\approx1$. Furthermore, the wave function transforms as an $A_2$ irreducible representation of the $D_{3}$ point group, invariant under $C_{3z}$ and odd under $C_{2y}$, lattice version of an $f$-wave~\cite{Cr_pel_2021,Cr_pel_2022,Cr_pel_Cea_2022}. 
We quantify the spread through the mean square radius $\langle r^2\rangle$ defined as
\begin{equation}\label{mean_square_radius}
    \langle r^2\rangle=\frac{\sum_{\Delta r}|\Psi_{2e}(r+\Delta r,r)|^2\overline{\Delta r}^2}{\sum_{\Delta r}|\Psi_{2e}(r+\Delta r,r)|^2}
\end{equation}
where $\overline{\Delta r}$ is the distance from $r$ module $(L_1,L_2)$, i.e. it is invariant under a shift of $\Delta r\to \Delta r+L_{1/2}$ with $L_{1/2}$ dimension of the cluster setting the largest length scale resolved in the numerics. Fig.~\ref{fig:2e_wavefunction}b) illustrates the mean-square radius computed numerically. 

\begin{figure}
    \centering
    \includegraphics[width=0.55\linewidth]{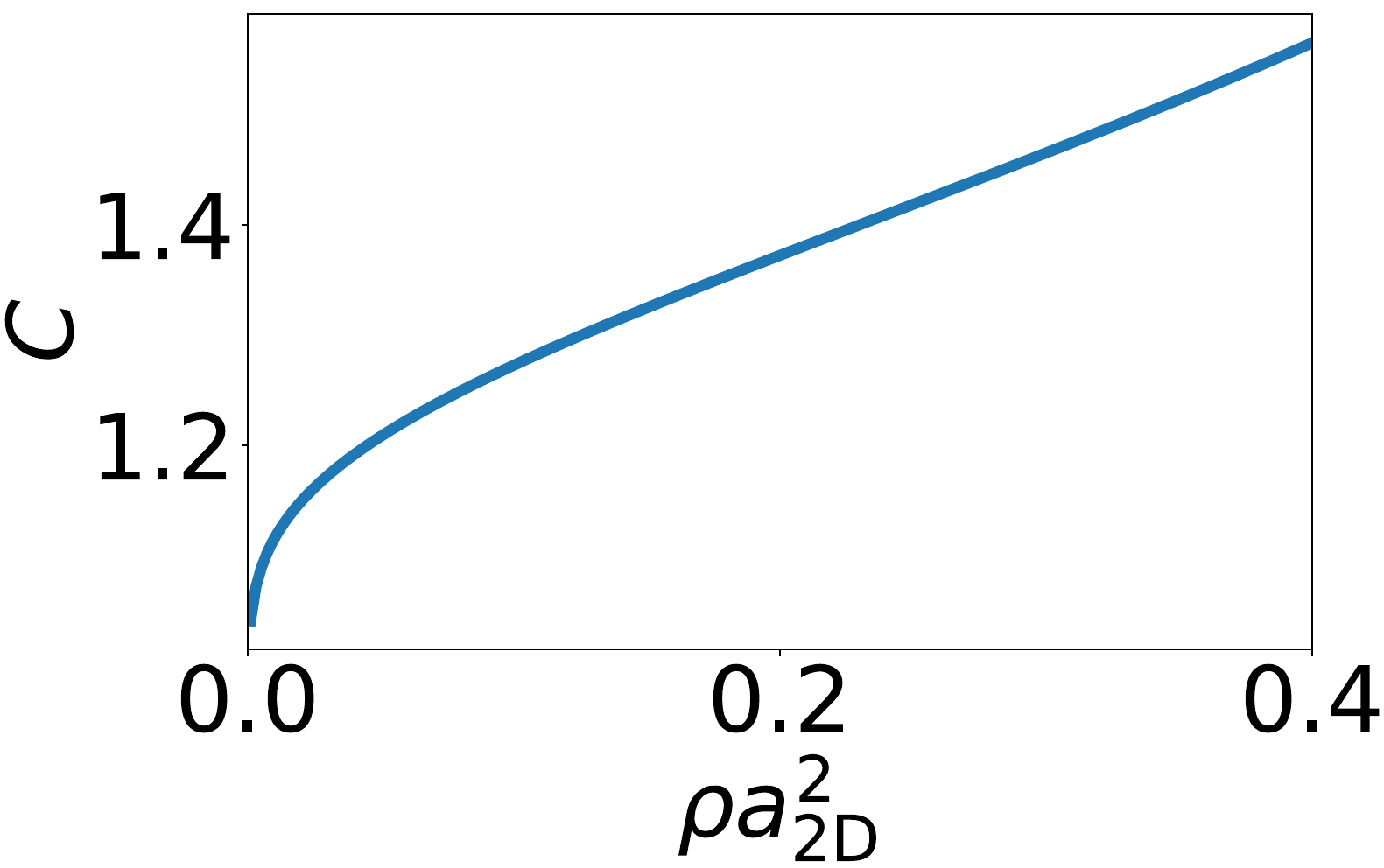}
    \caption{Evolution of the coefficient $C$ as a function of $\rho a^2_{\rm 2D}$.}
    \label{fig:sm_coefficient}
\end{figure}

\begin{figure}
    \centering
\includegraphics[width=0.5\linewidth]{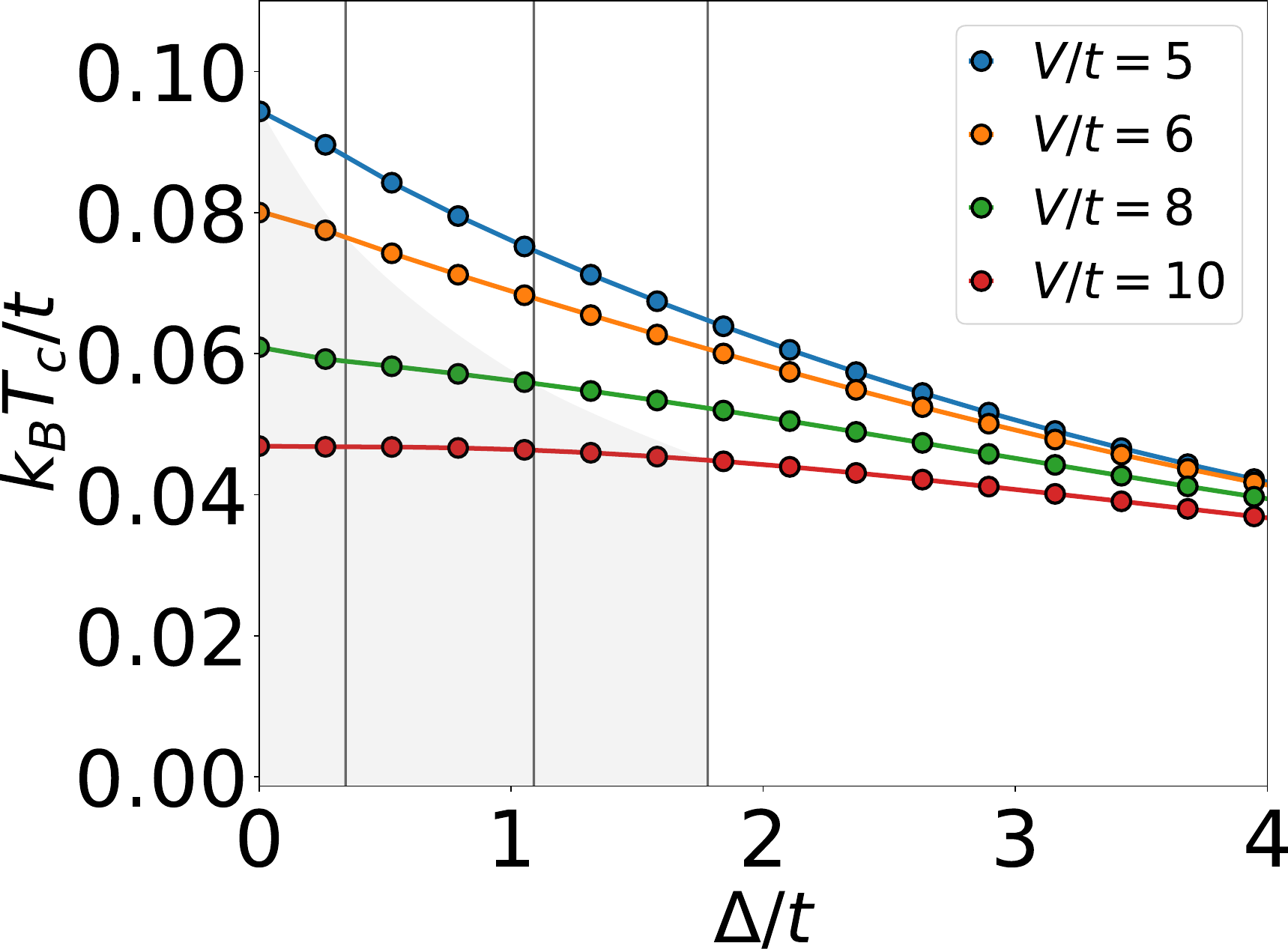}
    \caption{Optimal critical temperature computed from ED simulations. }
    \label{fig:critical_temperature}
\end{figure}

From the evaluation of the effective mass $m_b$ and the mean square radius $\langle r^2 \rangle$, we estimate the optimal critical temperature as~\cite{PhysRevB.37.4936,PhysRevLett.87.270402,PhysRevLett.100.140405,zhang_Tc_2023,szhang_prx_2023,sous_prb_2024}: 
\begin{equation}\label{sm:tc}
    T_c= C\frac{\hbar^2\rho}{m_b},    
\end{equation}
where $\rho=1/(\pi \langle r^2\rangle)$. In Eq.~\ref{sm:tc}, the coefficient $C$ depends on the repulsive interaction between the bosons. Specifically, we have~\cite{PhysRevB.37.4936,PhysRevLett.87.270402,PhysRevLett.100.140405,supplementary}:
\begin{equation}
    C=\frac{2\pi}{\log(380/4\pi)+\log\log(1/\rho a^2_{\rm 2D})},
\end{equation}
where $\rho$ is the density of bosons and $a_{\rm 2D}$ is the 2D scattering length. The evolution of $C$ is shown in Fig.~\ref{fig:sm_coefficient}, and due to the double logarithmic behavior, it exhibits only a weak dependence on $\rho a^2_{\rm 2D}$. We set $C\approx2\pi/\log(380/4\pi)$ in our calculations.
Fig.~\ref{fig:critical_temperature} shows the evolution of $T_c$ as a function of $\Delta$ for different values of $V/t$. The shaded gray area highlights the region where an attraction between two excitonic Cooper pairs emerges. In this regime, at finite density, the system's properties are governed by a complex interplay between charge density waves and superconductivity, which will be explored in future studies.

\section{Exact results in the $V\to\infty$ limit}
\label{app:infinite_coupling}

In this section we present exact results on the spectrum of charge excitations of the model. The $V\to\infty$ constraint selects only intersublattice excitations preserving the number of nearest-neighbor intersublattice pairs. As a result, in this limit, charged excitations are localized and the exact spectrum of eigenstates is found by diagonalizing the Hamiltonian on a finite size cluster.

\subsection{charge-$2e$ excitation}

\begin{figure}
    \centering
    \includegraphics[width=0.65\linewidth]{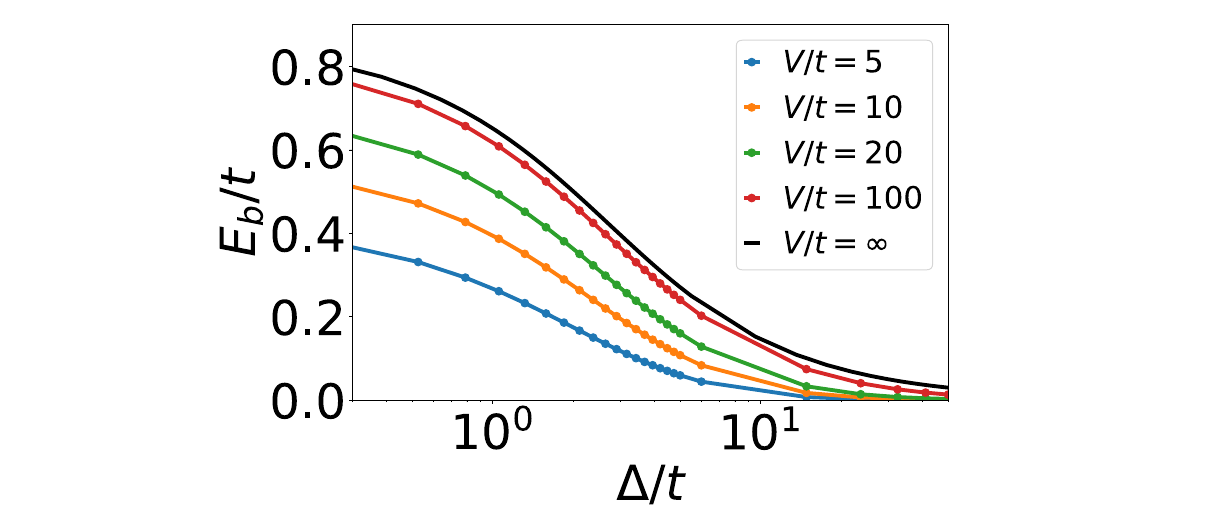}
    \caption{Binding energy as a function of $\Delta/t$ for different $V/t$. The solid black line shows the binding energy for  $V\to\infty$.}
    \label{fig:binding_sm}
\end{figure}

For strong repulsion, a pair of doped carriers is confined to a triangular cluster and  exhibits a finite binding energy, making the charge-$2e$ complex energetically favorable compared to two separate charge-$e$ excitations. 
The binding energy originates from the charge-transfer exciton connecting the $2e$ state with the trimer configuration. The Hamiltonian describing the four-sites cluster reads: 
\begin{equation}
    H_{2e} = \Delta \sum_{j=1}^{3} (f^\dagger_j f_j-2)-t \sum_{j=1}^{3}\left(f^\dagger_j f_{0} + h.c.\right),
\end{equation}
with four sites and total charge $Q=3$. We readily find two zero energy configurations: 
\begin{equation}
    \ket{E_1}=\frac{f^\dagger_1f^\dagger_2-f^\dagger_2f^\dagger_3}{\sqrt{2}}f_0^\dagger\ket{0},\quad \ket{E_2}=\frac{f^\dagger_1f^\dagger_2+f^\dagger_1f^\dagger_3}{\sqrt{2}}f_0^\dagger\ket{0},
\end{equation}
forming a two-dimensional irreducible representation, $C_{3z}\ket{E_1}=\ket{E_2}-\ket{E_1}$ and $C_{3z}\ket{E_2}=-\ket{E_1}$. In addition, we find two $C_{3z}$ invariant configurations $\prod_{j=1}^{3}f^\dagger_j\ket{0}$ and $\left(\sum_{j=1}^{3}f^\dagger_j f^\dagger_{j+1}\right)f^\dagger_0\ket{0}/\sqrt{3}$. Projecting $H_{2e}$ in this two-dimensional subspace we find that the ground state of the model is the bonding configuration: 
\begin{equation}\label{groundstate_4sites}\begin{split}
    &\ket{A} = \sqrt{\frac{1}{2}-\frac{\Delta}{4\sqrt{\Delta^2/4+3t^2}}} \prod_{j=1}^{3}f^\dagger_j\ket{0} + \sqrt{\frac{1}{2}+\frac{\Delta}{4\sqrt{\Delta^2/4+3t^2}}}\left(\frac{1}{\sqrt{3}}\sum_{j=1}^{3}f^\dagger_j f^\dagger_{j+1}\right)f^\dagger_0\ket{0},\\
    &E_{A} =\frac \Delta 2 -\sqrt{\frac{\Delta^2}{4}+3t^2},
\end{split}\end{equation}
which belongs to the $A_2$ irreducible representation of the point group of the crystal. 
We conclude that the binding energy reads: 
\begin{equation}\label{bbbinding_sm}
    E_{b} = E_{1e}-E_{2e}/2=\frac{1}{2}\sqrt{\frac{\Delta^2}{4}+3t^2}-\frac \Delta 4  \ge 0.
\end{equation}
We emphasize that trimer charge fluctuations serve as the glue that enables the formation of the bound state. 
Before proceeding, we present the evolution of the binding energy as a function of $\Delta/t$ for various values of $V/t$ in Fig.~\ref{fig:binding_sm}. Increasing $V/t$ the results obtained with ED approaches the asymptotic value in Eq.~\eqref{bbbinding_sm}.

Before concluding this section, we analyze the symmetries of the excitonic Cooper pair. To this end, we consider the charge-$2e$ wavefunction $\Psi_{2e}(r+\Delta r,r)$~\eqref{2ewavefunction} in the $V\to\infty$ limit, where we obtain the exact expression: 
\begin{equation}
    \Psi_{2e}(r+\Delta r,r)=\mel{\Phi_0}{f_{r+\Delta r}f_r}{\Phi_2}=\delta_{\Delta r,u_i-u_j}\sin 3\theta_{ij},
\end{equation}
with $r\in B$ and $\theta_{ij}=\arg(u_i-u_j)$ and the $2e$ bound state is confined within a single unit cell. Under a $C_{3z}$ rotation around $r$, the wavefunction remains invariant, $C_{3z}\Psi_{2e}=\Psi_{2e}$, this follows from $C_{3z}\theta_{ij}=\theta_{ij}+2\pi/3$. Moreover, $\Psi_{2e}$ changes sign under $C_{2y}$ rotations, i.e., $C_{2y}\Psi_{2e} = -\Psi_{2e}$, because $C_{2y}\theta_{ij}=\theta_{ij}+\pi$.  
Consequently, $\Psi_{2e}$ belongs to the $A_2$ irrep of $D_3$, lattice version of a $f$-wave~\cite{supplementary}. The property persists for finite value of $V/t$, as displayed in Fig.~\ref{fig:2e_wavefunction} by the numerical evaluation of $\Psi_{2e}$. Finally, associated to the charge-$2e$ bound state~\eqref{groundstate_4sites} we introduce a new emergent quasiparticle $b^\dagger_r$ located at $r$ with bosonic statistics:
\begin{equation}\label{single_boson}
    b^\dagger_r=\sqrt{1-|\alpha|^2}\prod_{j=1}^{3} f^\dagger_{r'_j}f_r+\frac{\alpha}{\sqrt{3}}\sum^3_{j=1}f^\dagger_{r'_j}f^\dagger_{r'_{j+1}},
\end{equation}
where $r'_j$ ($j=1,2,3$) are the lattice sites $r+u_j$, which are nearest neighbors of site $r$. Acting on the sublattice-polarized insulator $\ket{\Phi_0}$ generates the configuration $\ket{\Phi_2(r)} = b^\dagger_r \ket{\Phi_0}$.

\subsection{charge-$4e$ excitation}

We now calculate the energy of a charge-$4e$ composed by two charge-$2e$ pairs considering the case where the Cooper pairs are centered around neighboring sites and next to nearest-neighbor sites.

\subsubsection{Nearest-Neighbor Repulsion} 

\begin{figure}
    \centering
\includegraphics[width=0.5\linewidth]{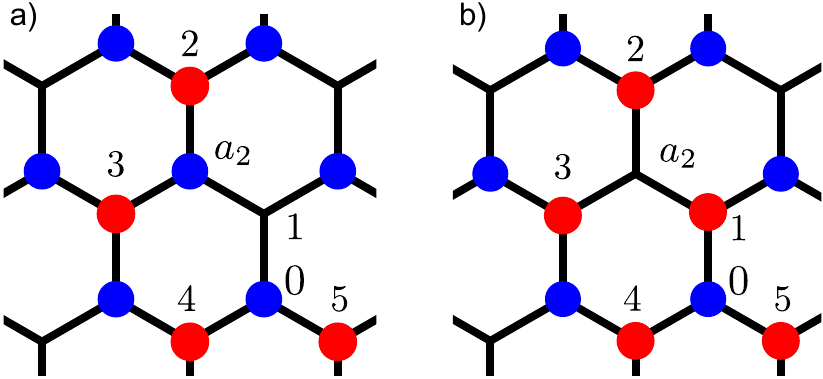}
    \caption{Panel a) shows one of the 5 configurations corresponding to adding 4 carriers in 5 sites $j=1,\cdots,5$. Panel b) shows the configuration obtained exciting a trimer centered around $a_2$.}
    \label{fig:two_pairs_nn}
\end{figure}
We start considering two Cooper pairs centered around $r=0$ and $r=a_2$ in Fig.~\ref{fig:two_pairs_nn} and showing that as a result of the Pauli exclusion principle we find a repulsive interaction between the two.  
We have $5$ configurations with 4 carriers in the $B$-sublattice: 
\begin{equation}
    \ket{\{j\}} = f^\dagger_{j_1}f^\dagger_{j_2}f^\dagger_{j_3}f^\dagger_{j_4}f^\dagger_0f^\dagger_{a_1}\ket{0},
\end{equation}
with $\{j\}$ denotes the five independent ways to select four labels from a set of five sites. Additionally, we have two excited states corresponding to a single trimer excitation centered around $a_2$ and $0$, respectively, with energy $\Delta$: 
\begin{equation}
    \ket{a_2}=\prod_{j=1}^{5} f^\dagger_jf^\dagger_0\ket{0},\quad \ket{0} = \prod_{j=1}^{5} f^\dagger_jf^\dagger_{a_2}\ket{0}.
\end{equation}
Computing the overlaps between the different configurations we find the Hamiltonian on nearest-neighbors: 
\begin{equation}
    H_{4e}=\begin{pmatrix}
        0 & 0 & 0 & 0 & 0 & -t & t \\
        0 & 0 & 0 & 0 & 0 & 0 & -t \\
        0 & 0 & 0 & 0 & 0 & 0 & t \\
        0 & 0 & 0 & 0 & 0 & t & 0 \\
        0 & 0 & 0 & 0 & 0 & -t & 0 \\
        -t & 0 & 0 & t & -t & \Delta & 0 \\
        t & -t & t & 0 & 0 & 0 & \Delta 
    \end{pmatrix}.
\end{equation}
The ground state energy is $E_{gs}=\Delta/2-\sqrt{8t^2+\Delta^2}/2>2E_{A}$~\eqref{groundstate_4sites} implying a net repulsive interaction between Cooper pairs on n.n. sites.
 
\subsubsection{Next to Nearest-Neighbor Attraction} 

\begin{figure}
    \centering
    \includegraphics[width=0.4\linewidth]{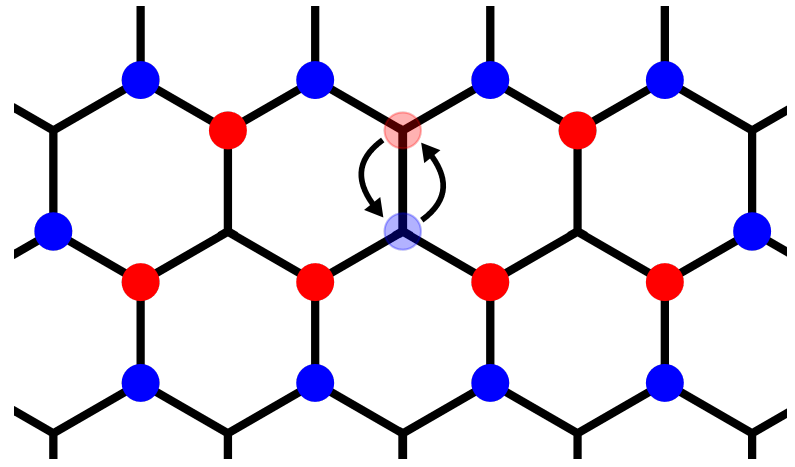}
    \caption{Two Cooper pairs centered around two next-n.n. sites, i.e. positioned at $r$ and $r+2a_1$, lower their energy by a virtual exciton involving the  intermediate center at $r+a_1$. }
    \label{fig:4e_resonance}
\end{figure}

We consider two Cooper pairs, one positioned at $r=0$ and the other at $r=2a_1$. Employing the bosonic operator $b^\dagger_r$~\eqref{single_boson} the corresponding configuration is: 
\begin{equation}
    \ket{\Phi_4(r,r+2a_1)}=b^\dagger_{r}b^\dagger_{r+2a_1}\ket{\Phi_0}.
\end{equation}
The Coulomb restricted tunneling $T_0=-t\mathbb P_{12}\sum_{\langle r,r'\rangle}f^\dagger_r f_{r'}\mathbb P_{12}$ with $12$ counting the number of n.n. bonds introduces quantum dynamics in the low-energy manifold which lowers the energy of the $4e$ charge complex with respect to the one of two isolated Cooper pairs. One among many processes is drawn in Fig.~\ref{fig:4e_resonance} where the middle site between the two Cooper pairs is resonating along $u_1$. We anticipate that in the infinite $V$ regime the Cooper pairs form a crystal (bosonic CDW) on n.n.n. sites with enlarged unit cell $(3a_1,2a_2)$. 
\begin{figure}
    \centering
    \includegraphics[width=0.5\linewidth]{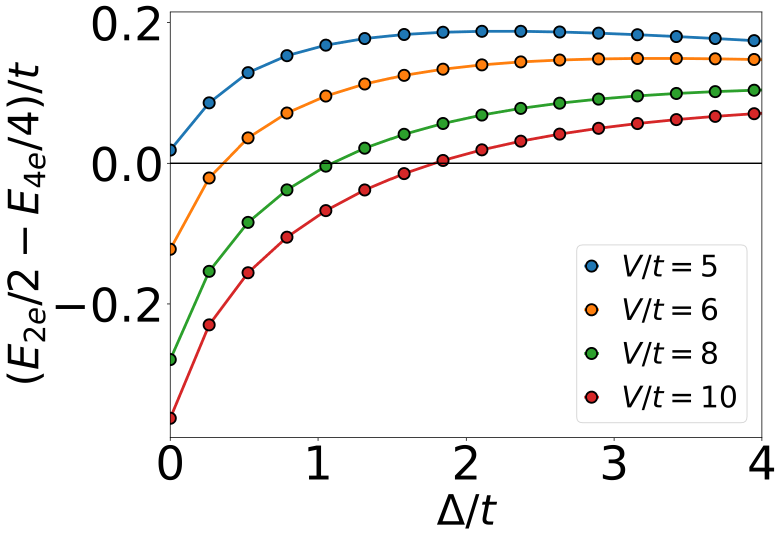}
    \caption{Energy difference $E_{2e}/2-E_{4e}/4$ indicating a tendency to form a four-body charge complex when $E_{2e}/2-E_{4e}/4<0$ for different values of $V/t$. }
    \label{fig:4e_bound_state}
\end{figure}

Fig.~\ref{fig:4e_bound_state} show ED simulations at finite $V/t$ we find that the repulsion between excitonic Cooper pairs is repulsive for $V/t<5$. For larger interaction strength, the boson $b_r$ develops a net attractive interaction leading to the formation of a four-body bound state. In this regime, increasing $\Delta/t$ tunes the interaction between Cooper pairs from attractive to repulsive.

\subsection{Infinite coupling model}

\begin{figure}
    \centering
    \includegraphics[width=0.5\linewidth]{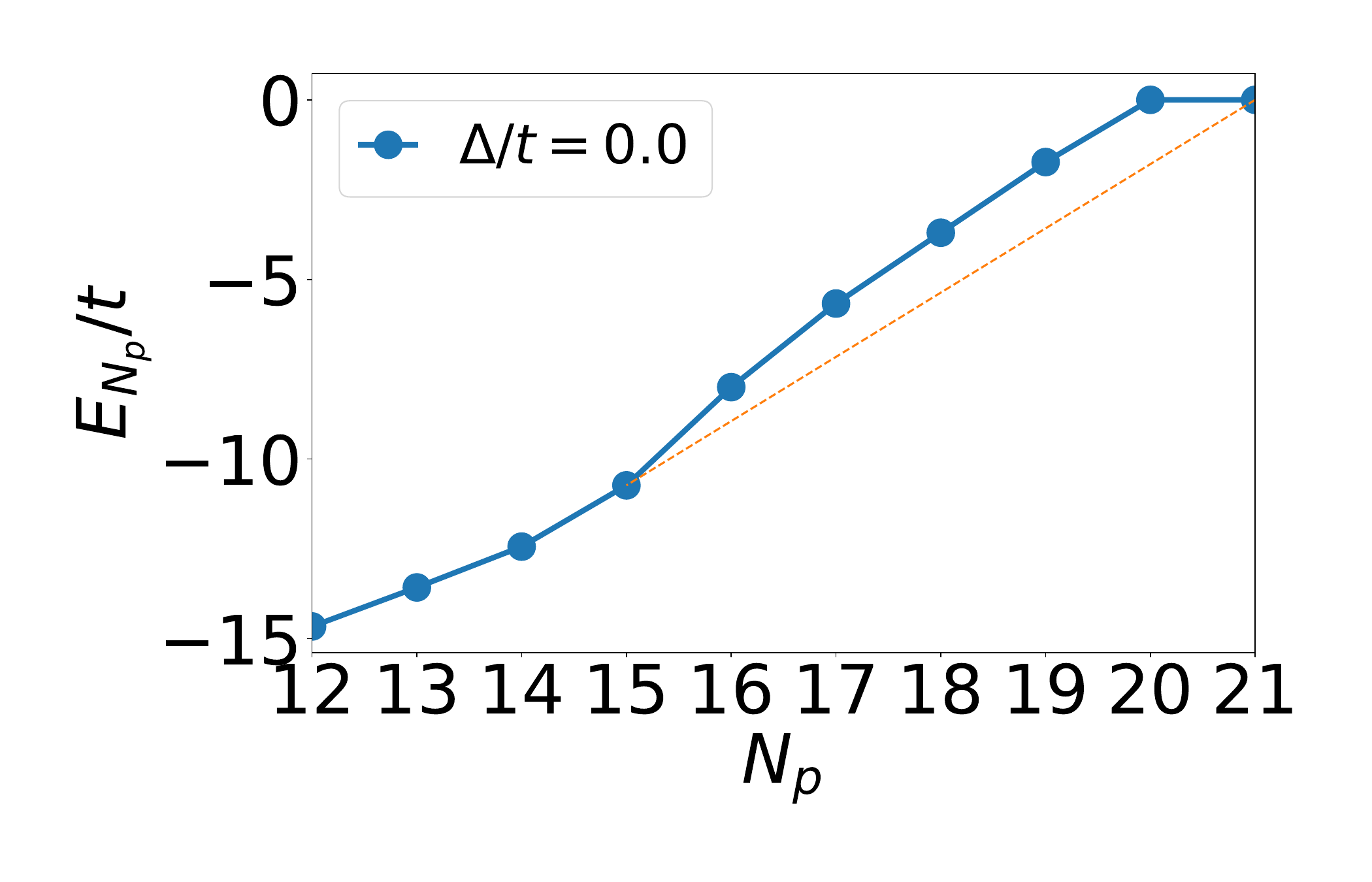}
    \caption{Ground state energy as a function of the number of particles $N_p$ obtained performing ED calculations on a cluster of 42 sites.}
    \label{fig:fig_phase_separation}
\end{figure}

In this section, we discuss the ground state properties obtained for larger doping in the limit $V\to\infty$. 
Exploiting the particle-hole symmetry of the model, we focus on the regime $N_p<N$ (hole-doped) with $N$ number of unit cells. 
In this regime, the Hamiltonian~\eqref{hamiltonian_minimal} takes the simple form: 
\begin{equation}
    H_\infty = -t\sum_{\langle r,r'\rangle } P_r f^\dagger_{r} f_{r'}P_{r'} + \Delta N_B,
\end{equation}
where hopping is permitted only between configurations that avoid nearest-neighbor occupancies:
\begin{equation}
    P_r = \prod_{r'\, \text{next to}\, r }(1-n_{r'}).
\end{equation}
The model features a single dimensionless parameter $\Delta/t$ and the constraint in the hopping makes the problem strongly interacting. 
Fig.~\ref{fig:fig_phase_separation} shows the ground state energy for different numbers of carriers in the hole-doped regime. The concave behavior observed in the energy curve between $15$ and $21$ provides evidence for a tendency toward phase separation within this doping range.

\section{Strong coupling perturbation theory}
\label{app:strong_coupling}

Here, we present the first-order correction in $t/V$ for the large $V$ regime. We discuss the corrections to the ground state energy for filling factors $N$, $N+1$ and $N+2$, binding energy and effective mass. 
Perturbation theory in small $t/V$ is performed organizing the spectrum of the model in sectors with a fixed number n.n. occupied sites $M$. 
The Hamiltonian decomposes as $H=H_0+H'$ with $H_{0}$ block diagonal in the subspace $M$, i.e. it does not mix different subspaces: 
\begin{equation}\label{diagonal}
    H_{0}=\sum_M H_M,\quad  H_{M}= \Delta N_B +MV.
\end{equation}
The off-diagonal contribution instead mixes the subspace $M$ with $M+q$:  
\begin{equation}\label{offdiagonal}
    H'=\sum_M\sum_{q}T_{q,M},\quad T_{q,M}=-t\mathbb P_{M+q}\sum_{\langle r,r'\rangle }f^\dagger_r f_{r'}\mathbb P_M,
\end{equation}
where $q=\pm1,\pm2$ corresponds to the addition/removal of $q$ n.n. occupied sites. 
To connect with previous works~\cite{Slagle_prb,Cr_pel_2021}, we observe that the action of $T_1$ creates a polaron and $T_2$ creates a dipole. 
Finally, we emphasize that $T_{0,M}$ preserves $\sum_{\langle r,r'\rangle} n_rn_{r'}$ which is equal to $MV$ but does not commute with $N_A-N_B$, thereby introducing quantum dynamics within each sector $M$.

\subsection{Schrieffer-Wolff transformation: effective Hamiltonian upon doping the sublattice polarized state}

In this section, we clarify the connection between the many-body perturbation theory, and the Schrieffer-Wolff transformation in the many-body space. 
Without losing generality, we consider charge sector of $N+p$ particles with $N$ number of sites and $p$ extra doped carriers. Moreover, for simplicity we consider the case $\Delta=0$, where for large $V$ the ground state spontaneously polarize in one of the two sublattices. 
Within each sector, we introduce the basis of states $\ket{\Phi_{p,nM}}$ where $p$ denotes the sector with $p$ extra doped carriers $(p=0,1,2)$, $M$ the number of n.n. occupied sites and $n$ the principal quantum number. By definition the unperturbed Hamiltonian projected in the particle sector $p$ reads: 
\begin{eqnarray}
    H_0=\sum_M\sum_n E_{nM} \ket{\Phi_{p,nM}}\bra{\Phi_{p,nM}},
\end{eqnarray}
 while $H'$ is off-diagonal and has the form:
 \begin{equation}
     H'=\sum_M\sum_{q}\sum_{nm} (T_{q,M})_{nm}\ket{\Phi_{p,nM+q}}\bra{\Phi_{p,mM}}.
 \end{equation}
By definition we have: 
\begin{equation}\label{commutator1}
    [H_0,T_{q,M}]=qV T_{q,M}.
\end{equation}
Moreover, we have the relations: 
\begin{equation}\label{commutator2}\begin{split}
    &[H_0,T_{q_1,M_1}T_{q_2,M_2}]=(q_1+q_2)V T_{q_1,M_1} T_{q_2,M_2}, \\
    &[H_0,[T_{q_1,M_1},T_{q_2,M_2}]]=(q_1+q_2)V [T_{q_1,M_1},T_{q_2,M_2}],
\end{split}\end{equation}
which directly follows from the properties of the commutator. 
Notice that for any particle sector $p$ the lowest energy configuration has a number of n.n. occupied sites $M=zp$ with $z=3$ coordination of the lattice. Moreover, for $p=0$ the ground state is only two-fold degenerate corresponding to a fully polarized sublattice insulator. On the other hand, for $p=1,2$ the ground state is extensively degenerate.

The Schrieffer-Wolff transformation consists of introducing the antiunitary operator $S$, $S^\dagger=-S$, leading to the transformed Hamiltonian: 
 \begin{equation}
     \bar H = e^{S}He^{-S}.
 \end{equation}
Employing Baker-Campbell-Hausdorff identity to third order in $S$, we have: 
 \begin{equation}\label{effective_Hamiltonian}
 \begin{split}
     \bar H &= H + [S,H]+\frac{1}{2}[S,[S,H]]+\frac{1}{3!}[S,[S,[S,H]]]+\cdots.\\ 
\end{split}
 \end{equation}

Our task is to define $S$ which order by order in $(t/V)$ allows to remove terms off-diagonal in the number of n.n. occupied sites $M$. Specifically, the transformation $S^{(k)}\sim\mathcal O[(t/V)^k]$ removes all off-diagonal terms of order $(t/V)^{k-1}$. To start with we define, the Hamiltonian at stage $k=1$ as: 
\begin{eqnarray}
    \bar H^{(1)} \equiv H=H_0+H',
\end{eqnarray}
and $S^{(0)}=1$ is the identity. At stage $k=2$, we have:  
\begin{equation}
    \bar H^{(2)} = e^{S^{(1)}} H e^{-S^{(1)}}=H+[S^{(1)},H_0] + [S^{(1)},H'] + \frac{1}{2}[S^{(1)},[S^{(1)},H_0]]+ \mathcal O\left(\frac{t^3}{V^2}\right).
\end{equation}
Imposing the conditions $[S^{(1)},H_0]=-\sum_{q\neq 0}\sum_M T_{q,M}$, we find: 
 \begin{equation}
     S^{(1)}=\sum_M\sum_{q\neq 0}\sum_{nm} \frac{(T_{q,M})_{nm}}{qV}\ket{\Phi_{p,nM+q}}\bra{\Phi_{p,nM}},
 \end{equation}
where we have introduced the notation $(T_{q,M})_{nm}\equiv \mel{\Phi_{p,nM+q}}{T_{q,M}}{\Phi_{p,mM}}$. As a result, the second order Hamiltonian $H^{(2)}$ reads:
\begin{equation}
    \bar H^{(2)} = H_0 + \sum_{M} T_{0,M}+ \sum_{M_1M_2}\sum_{q\neq 0}\frac{[T_{q,M_1},T_{0,M_2}]}{qV}+\frac{1}{2} \sum_{M_1M_2}\sum_{q_1q_2\neq 0}\frac{[T_{q_1,M_1},T_{q_2,M_2}]}{q_1V}+\mathcal O\left(\frac{t^3}{V^2}\right).
\end{equation}
Projecting the Hamiltonian in the lowest energy manifold with number of n.n. occupied sites $M=zp$ we obtain: 
\begin{equation}
    \mathcal H^{(p)}=H_{zp}-\sum_{q=1}^{2}\frac{T^\dagger_{q,zp}T_{q,zp}}{qV}+\mathcal O\left(\frac{t^3}{V^2}\right).
\end{equation}
where $H_{zp}=H_0+T_{0,zp}$. 

\subsubsection{Next-to-leading order corrections}

To obtain an accurate expression for the effective mass of the charge-$2e$ excitation and the binding energy valid to larger values of $t/V$, we extend our perturbation theory up to order $t^3/V^2$. To this aim we introduce $S^{[2]}\sim (t/V)^2$ which removes off-diagonal terms of order $t/V$: 
\begin{equation}\begin{split}
 \bar H^{(3)} =& e^{ S^{(2)}} H e^{-S^{(2)}} = H_0 + \sum_{M} T_{0,M}+ \sum_{M_1M_2}\sum_{q\neq 0}\frac{[T_{q_1,M_1},T_{0,M_2}]}{q_1V}+\frac{1}{2} \sum_{M_1M_2}\sum_{q_1q_2\neq 0}\frac{[T_{q_1,M_1},T_{q_2,M_2}]}{q_1V}\\
 &+[S^{[2]},H_0]+[S^{[2]},H']+ \sum_M\frac{[S^{(1)},[S^{(1)},T_{0,M}]]}{2} +  \sum_{q\neq 0}\sum_M\frac{[S^{(1)},[S^{(1)},T_{q,M}]]}{3},   
\end{split}\end{equation}
 where $S^{(2)}=S^{[1]}+S^{[2]}$ with $S^{[1]}=S^{(1)}$ and $S^{[2]}$ such that: 
 \begin{equation}
     [S^{[2]},H_0] = -\sum_{M_1M_2}\sum_{q_1\neq 0}\frac{[T_{q_1,M_1},T_{0,M_2}]}{q_1V}-\frac{1}{2} \sum^{M_1+q_1\neq M_2}_{M_1M_2}\sum_{q_1q_2\neq 0}\frac{[T_{q_1,M_1},T_{q_2,M_2}]}{q_1V}.
 \end{equation}
 $S^{[2]}$ is obtained employing the identity~\eqref{commutator1} and is composed by the sum of two contributions: 
\begin{equation}
    S^{[2]}=\sum_{M_1M_2}\sum_{q_1\neq 0}\frac{[T_{q_1,M_1},T_{0,M_2}]}{(q_1V)^2}+\frac{1}{2}\sum^{M_1+q_1\neq M_2}_{M_1M_2}\sum_{q_1q_2\neq 0}\frac{[T_{q_1,M_1},T_{q_2,M_2}]}{q_1(q_1+q_2)V^2}.
\end{equation}

 The resulting third order Hamiltonian reads: 
 \begin{equation}\begin{split}
 \bar H^{(3)} =& H_0 + \sum_{M} T_{0,M}+\frac{1}{2} \sum^{M_1+q_1=M_2}_{M_1M_2}\sum_{q_1q_2\neq 0}\frac{[T_{q_1,M_1},T_{q_2,M_2}]}{q_1V}\\
 &+[S^{(2)},H']+ \sum_M\frac{[S^{(1)},[S^{(1)},T_{0,M}]]}{2} +  \sum_{q\neq 0}\sum_M\frac{[S^{(1)},[S^{(1)},T_{q,M}]]}{3}+\mathcal O\left(\frac{t^4}{V^3}\right),  
\end{split}\end{equation}
which is diagonal up to order $t/V$. The effective Hamiltonian including corrections to order $t^3/V^2$ is then obtained projecting $\bar H^{(3)}$ in the low-energy manifold with $M=zp$ n.n. occupied sites. By performing, long but straightforward calculations we conclude:
\begin{eqnarray}
    \mathcal H^{(p)} = H_{zp} -\sum_{q=1,2}\frac{T^\dagger_{q,zp}T_{q,zp}}{qV}+\sum_{q=1,2}\frac{T^\dagger_{q,zp}T_{0,zp+q}T_{q,zp}}{(qV)^2}-\frac{1}{2}\sum_{q=1}^2\frac{\{T^\dagger_{q,pz}T_{q,pz},T_{0,pz}\}}{(qV)^2},
\end{eqnarray}
where $H_{zp}=H_0+T_{0,zp}$ valid up to $\mathcal O(t^4/V^3)$.

\subsubsection{Filling $N_e=N_s$ ($\nu=1$)}

\begin{figure}
    \centering
    \includegraphics[width=0.7\linewidth]{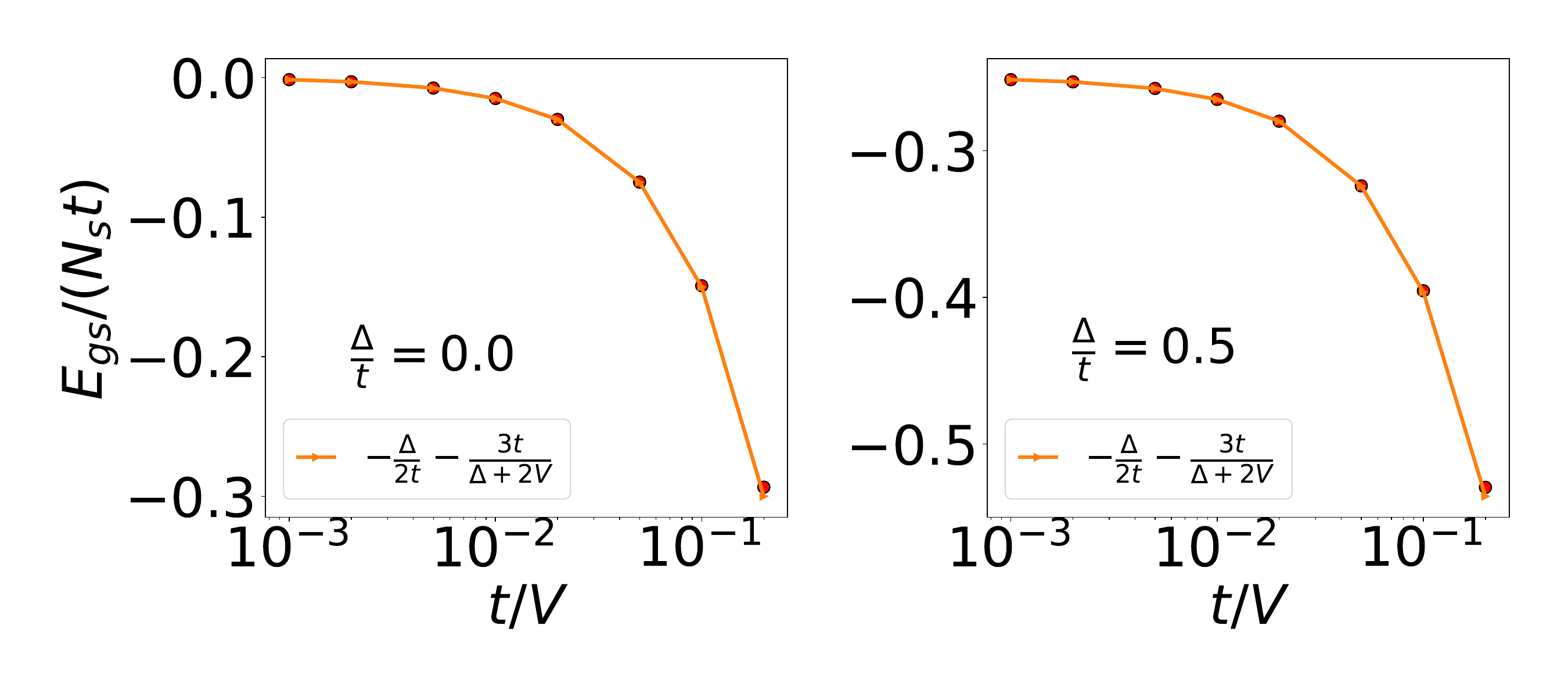}
    \caption{Left and right panel shows the evolution of the many-body ground state energy per particle as a function of $t/V$ obtained setting $\Delta=0$ in the left panel and $\Delta/t=0.5$ in the right one. Perturbation theory provides a reliable estimate of the ground state energy per particle, even for $V/t=5$ where the relative error is approximately $2\%$. }
    \label{fig:gs_energy}
\end{figure}

In the $V\to\infty$ limit, the ground state is determined by minimizing the Coulomb energy, $V\sum_{\langle r,r'\rangle} n_r n_{r'}$. This corresponds to the sublattice polarized insulator:
\begin{equation}
    \ket{\Phi_0}=\prod_{r\in A} f^\dagger_r\ket{0},
\end{equation}
belonging to the zero momentum sector $Q=\Gamma$. 
The first order correction in $t/V$ to the ground state energy reads: 
\begin{equation}
    \delta E_{0}= \sum_{n}\frac{\mel{\Phi_0}{T_{-2}}{\Phi_{0,n2}}\mel{\Phi_{0,n2}}{T_2}{\Phi_0}}{E^{(0)}_{0}-E^{(0)}_{0,n2}},
\end{equation}
where $\ket{\Phi_{0,\gamma,2}}$ is obtained by acting with $T_2$ on the fully sublattice polarized ground state $\ket{\Phi_0}$. 
Thus, the excited configuration corresponds to a dipole, with energy $\Delta + 2V$, situated along the bond connecting $r$ and $r + u_j$. In this case, the excited subspace is localized and the projected hopping $-t\mathbb P_{2}\sum_{\langle r,r'\rangle}f^\dagger_rf_{r'}\mathbb P_{2}$ in the subspace of $2$ n.n. occupied sites at filling $\nu=1$ is trivial and does not introduce any quantum dynamics. The energy correction reads: 
\begin{equation}
     \delta E_{0} = -N \frac{3t^2}{\Delta +2V},
\end{equation}
where $3$ is the coordination number of the honeycomb lattice. Fig.~\ref{fig:gs_energy} shows the comparison between the $t/V$ perturbation theory result and the ground state energy obtained from ED.

\subsubsection{charge-$e$ excitation}

\begin{figure}
    \centering
    \includegraphics[width=.7\linewidth]{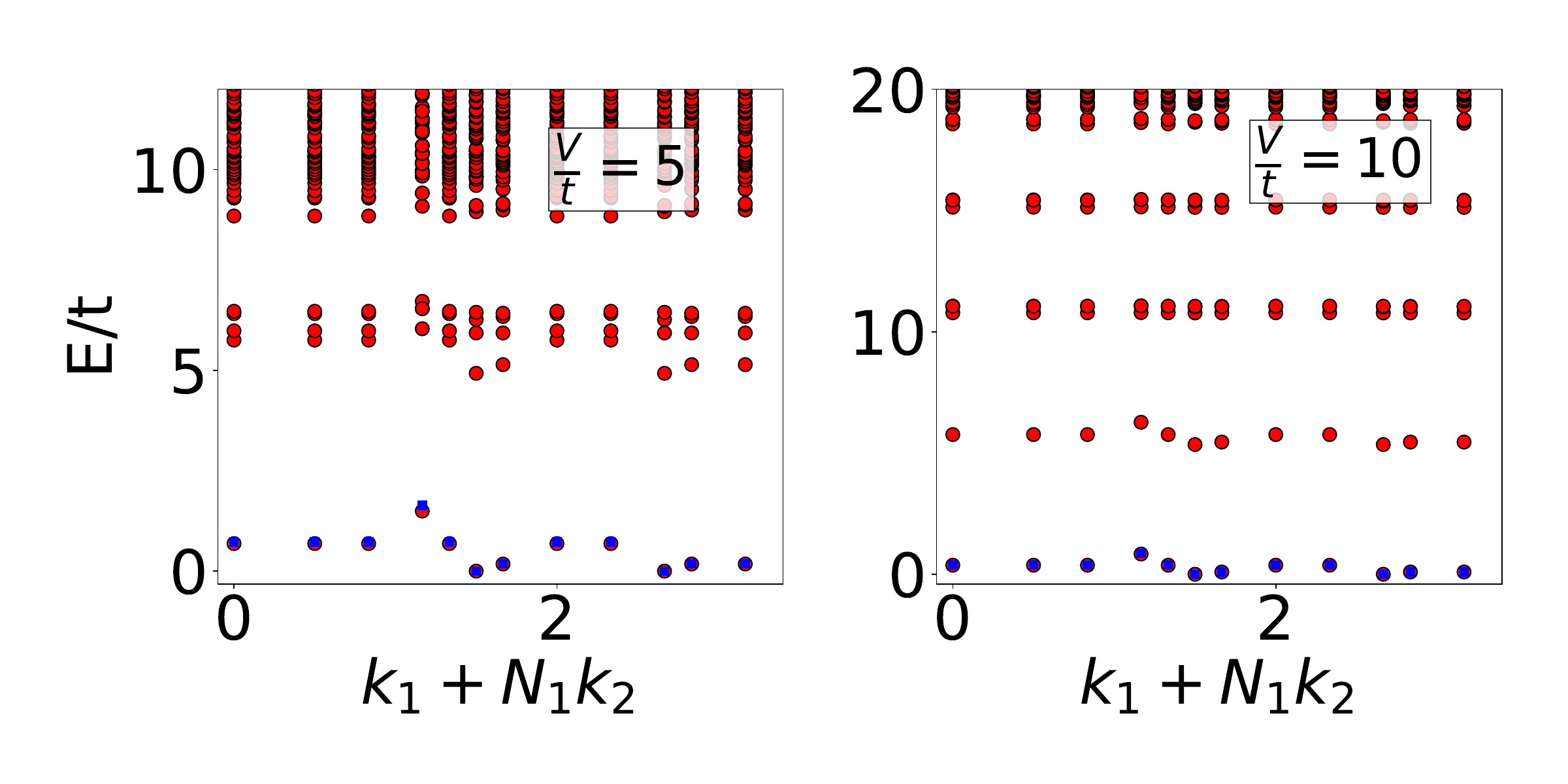}
    \caption{Many-body spectrum for the charge sector with $p=1$ doped charge excitation for $\Delta/t=0.5$. The lowest energy branch of the many-body spectrum describes the charge-$e$ excitation. Blue square data shows the analytical prediction for the dispersion relation obtained order $t/V$ in perturbation theory.}
    \label{fig:manybody_spectrum_1e}
\end{figure}

In the strong coupling limit $V \to \infty$, adding an extra carrier incurs an energy cost of $E_{1e} = \Delta + 3V$ with an infinite effective mass, thus, to leading order, realizing a charge-$e$ quasiparticle in a perfectly flat band.
The effective hopping of the charge-$e$ excitation involves an intermediate polaron excitation generated by $T_1$ and leads to the hopping Hamiltonian: 
\begin{equation}\label{sm:1e_ptheory}
    H_f=t_f\sum_{\langle r,r'\rangle\in B}f^\dagger_r f_{r'},\quad t_f =\frac{t^2}{\Delta+V}.
\end{equation}
The latter describes particles hopping on a triangular lattice with dispersion relation $\epsilon_k=2t_f\sum_{j=1}^{3}\cos(k\cdot a_j)$ displaying minima at $k=K,K'$ and effective mass $m_f=2/(3t_f)\sim V$. 

\begin{figure}
    \centering
    \includegraphics[width=.5\linewidth]{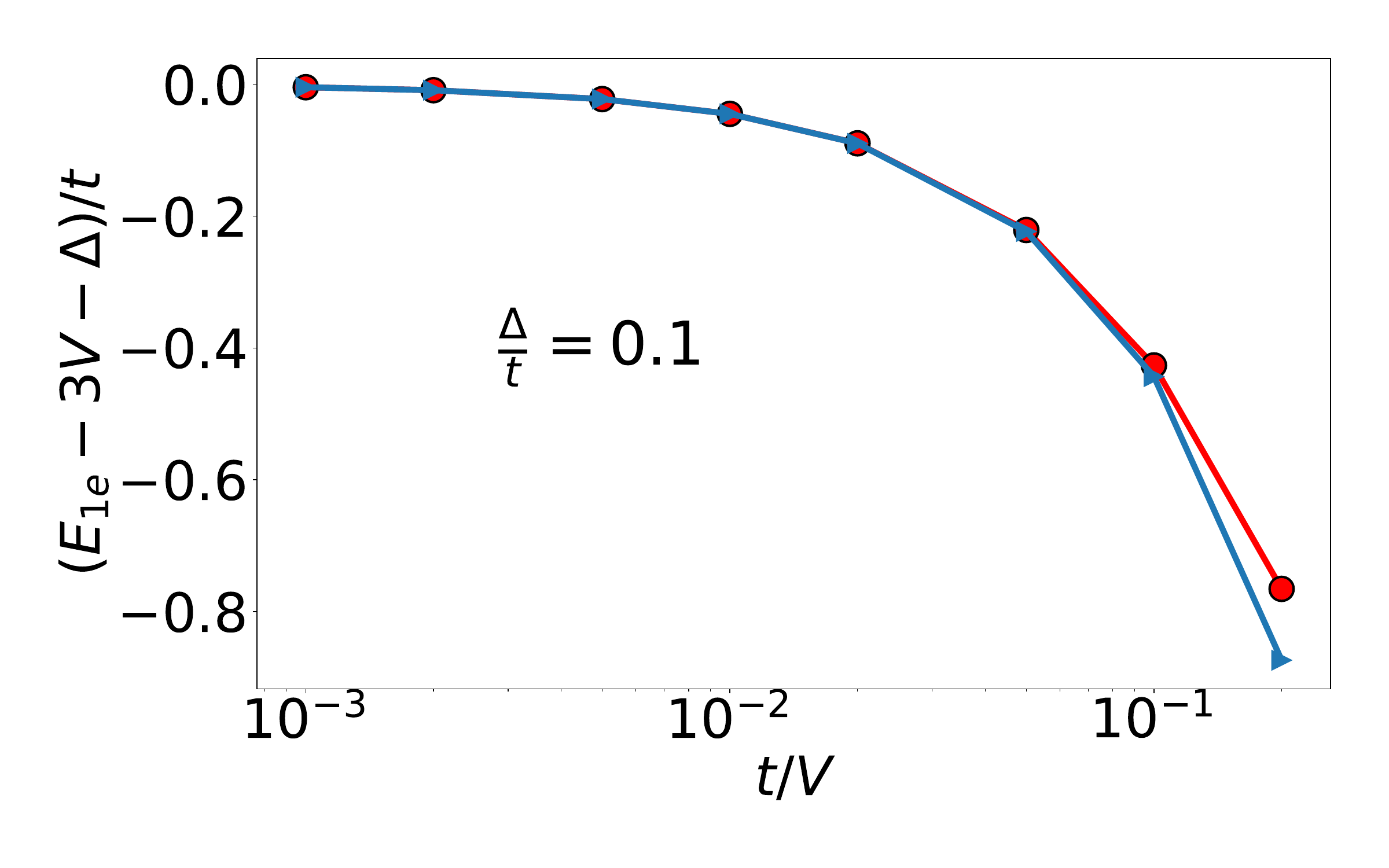}
    \caption{Energy of a charge-$e$ excitation measured with respect to the charging energy $\Delta+3V$. Reducing $V$ increases the difference with respect to the perturbative result. The maximum relative error is $14\%$ for $V/t=5$ and becomes negligible (smaller than $1\%$) for $V/t\ge20$.}
    \label{fig:Ns+1particles}
\end{figure}

We now derive the first order correction to the ground state energy in $t/V$. 
To this aim, we consider the wavefunction with total momentum $K$: 
\begin{equation}
    \ket{\Phi_1}=\frac{1}{\sqrt{N}}\sum_{r\in A}f^\dagger_{r+u_1} e^{-iK\cdot(r+u_1)}\ket{\Phi_0},
\end{equation}
where $K=(2g_1+g_2)/3$ and $g_{1/2}$ reciprocal lattice vectors. For each configuration $\ket{\Phi_1(r)}$, there are two different hopping processes: those involving the three sites $r$, $r+a_1$ and $r+a_6$ nearest-neighbors of the $B$-sublattice occupied site $r+u_1$ and the sites $r'$ away from this region. The second order energy correction originating from processes away from these sites is: 
\begin{equation}
    \delta E_{1,1}= -(N-3)\frac{3t^2}{\Delta+2V}.
\end{equation}
The contribution of the remaining sites $\{r,r+a_1,r+a_6\}$ is given by: 
\begin{equation}
    \delta E_{1,2} = -3\frac{2t^2}{\Delta+V}+\epsilon_{K/K'}=-\frac{9t^2}{\Delta+V}.
\end{equation}
Taking for simplicity the site $r$, the first contribution arises from the action of $-tf^\dagger_{r+u_{2/3}}f_r$, which creates a polaron that subsequently recombines without involving any motion of the additional doped carrier at $r$. Conversely, the second contribution represents the kinetic energy gain from adding the extra carrier at $K$. Notably, this correction can be explicitly determined by considering second-order processes that involve the motion of the doped carrier. The ground state energy to first order in $t/V$ reads: 
\begin{equation}
    E_{1} = \Delta + 3V  -(N-3)\frac{3t^2}{\Delta+2V} -\frac{9t^2}{\Delta+V}.
\end{equation}
Thus, we conclude that the lowest energy of a charge-$e$ quasiparticle up to first order in $t/V$ is:
\begin{equation}
    E_{1e} = E_{1}-E_{0}= \Delta + 3V -\frac{9t^2}{\Delta+V}+\frac{9t^2}{\Delta+2V}=\Delta + 3V -\frac{9t^2 V}{(\Delta+V)(\Delta+2V)}.
\end{equation}
Fig.~\ref{fig:Ns+1particles} shows the comparison between the energy of a single doped excitation obtained with ED (red data) and the result of perturbation theory both measured with respect to the charging energy $\Delta+3V$. 

\subsubsection{charge-$e$ polaron excitation}

In this section we characterize the properties of a charge-$e$ excitation coupled to a single dipole. This excited state has energy $\Delta+V$ above the charge-$e$ quasiparticle. Interestingly, the basis of many-body states spanning the manifold is composed by three different configurations classified by their different eigenvalues under $C_{3z}$:
\begin{equation}
    \begin{split}
        &\ket{\Phi_{1,\ell}(r)} = \frac{1}{\sqrt{3}}\sum_{j=1}^{3}\omega^{\ell(j-1)}f^\dagger_{r+u_j}f^\dagger_{r+u_{j+1}}f_{r}\ket{\Phi_0},\\
    \end{split}
\end{equation}
where $\omega=\exp(2\pi i/3)$ and $\ell=0,\pm$. These states form an orthonormal basis and features three different eigenvalues under $C_{3z}$.
Up to order $t/V$ the Hamiltonian in this subspace reads: 
\begin{equation}
    \mathcal H^{(1)}_{\rm polaron} = \Delta + V + \frac{T^\dagger_{-1,4}T_{-1,4}}{V+\Delta}-\sum_{q=1,2} \frac{T^\dagger_{q,4}T_{q,4}}{(qV+\Delta)},
\end{equation}
where the energy is measured with respect the charge-$e$ unperturbed energy $\Delta+3V$. The first contribution $q=-1$ corresponds to the process of recombination of the electron-hole excitation. 
This process does not induce any dispersion but produces an energy splitting, in the basis $[\ket{\Phi_{1,0}(r)},\ket{\Phi_{1,+}(r)},\ket{\Phi_{1,-}(r)}]^T$ we have:  
\begin{equation}
    \frac{T^\dagger_{-1,4}T_{-1,4}}{V+\Delta} = \frac{t^2}{V+\Delta}\begin{pmatrix}
        0 & 0 & 0 \\ 
        0 & 3 & 0 \\ 
        0 & 0 & 3
    \end{pmatrix}.
\end{equation}
On the other hand, $T^\dagger_{1,4}T_{1,4}/(V)$ induces hopping in real space. 
Before moving on, it is interesting to notice that: 
\begin{eqnarray}
    \mel{\Phi_{1,a}(r+a_j)}{T^\dagger_{1,4}T_{1,4}}{\Phi_{1,0}(r)}/V=0,\, \forall j,a
\end{eqnarray}
implying that the configuration is localized due to a destructive interference effect. 
On the other hand, $\ket{\Phi_{1,\pm}(r)}$ develops a finite dispersion relation. 
\begin{equation}
    \mathcal H^{(1)}_{\rm polaron}(k) =\Delta + V  +\frac{t^2}{\Delta+V}\begin{pmatrix}
        0 & 0 & 0 \\ 
        0 & 3+\sum_{j=1}^{3}\cos(k\cdot a_j+\pi/3) & \sum_{j=1}^{3}\cos k\cdot a_j \\ 
        0 & \sum_{j=1}^{3}\cos k\cdot a_j & 3+\sum_{j=1}^{3}\cos(k\cdot a_j-\pi/3)
    \end{pmatrix}.
\end{equation}
We found that the minimum of the dispersion is located at $K/K'$.

\subsubsection{Perturbative corrections to the charge gap} 

In this section, we discuss the renormalization of the charge gap due to perturbative corrections. 

To begin, we note that, to leading order in $t/V$, the energy of an isolated hole is $E_{-1}=0$. Including first order corrections, up to order $t^3/V^2$ we obtain: 
\begin{equation}
    E_{-1}=-(N-2)\frac{3t^2}{\Delta + 2V }-\frac{9t^2}{\Delta+V}.
\end{equation}
Finally, the charge gap reads: 
\begin{equation}
    E_{\rm gap}= E_1+E_{-1}-2E_0=\Delta +3V-3t^2\frac{\Delta+7V}{(\Delta+V)(\Delta+2V)}.
\end{equation}
Thus, perturbative corrections in $t/V$ reduce the charge gap size. Within perturbation theory the gap collapse at $V=1.87t$, overestimating the actual critical value of $1.3t$ displayed in Fig.~\ref{fig:filling_1}d).

\subsubsection{charge-$2e$ excitation}

In the $V \to \infty$ limit, a pair of $e$ excitations in a sublattice-polarized ground state binds to form an excitonic Cooper pair, represented by the bosonic quasiparticle $b_r$. 
As discussed previously, the state $\ket{\Phi_2(r)}=b^\dagger_r\ket{\Phi_0}$ is an eigenstate of $H_{M=6}$~\eqref{diagonal} with energy $E_A+2\Delta+6V$ and $E_A$ given in Eq.~\eqref{groundstate_4sites}. To leading order in $t/V$ has infinite effective mass resulting in a perfectly flatband of excitonic Cooper pairs. 
In the following we derive the effective mass, the ground state energy $E_{N+2}$ and the binding energy $E_{b}$ to first order in $t/V$.

\subsubsection{charge-$2e$ dispersion relation}

To order $t/V$ the effective hopping of the $2e$-excitation is determined performing degenerate perturbation theory in the low-energy manifold spanned by the basis of states $\ket{\Phi_2(r)}=b^\dagger_r\ket{\Phi_0}$ with $\braket{\Phi_2(r)}{\Phi_2(r')}=\delta_{r,r'}$. To leading order, the action of $H'$~\eqref{offdiagonal} induces an excitation with energy scaling as $qV$ (with $q = 1$ for a polaron excitation and $q=2$ for a dipole~\cite{Slagle_prb,Cr_pel_2021}). At a subsequent stage, this excitation recombines, connecting a Cooper pair initially located at $r$ to a Cooper pair at $r+a_j$. The resulting nearest-neighbor tunneling amplitude reads: 
\begin{equation}\label{sm:boson_tunneling}
    t_b =\sum_{q=1,2} \mel{\Phi_2(r+a_j)}{\left(T_{-q} \frac{1}{H_{q+6}-E^{(0)}_{2}} T_q\right)}{\Phi_2(r)} ,
\end{equation}
where $E^{(0)}_{2}$ is the unperturbed ground state energy for two doped carriers, $E^{(0)}_{2}=E_A+2\Delta+6V$. Furthermore, we observe that the amplitude in Eq.~\eqref{sm:boson_tunneling} does not depend on the nearest-neighbor site $r+a_{j=1,\cdots,6}$ as a result of the symmetries of the model. We note that the action of $T_{q}$ on $\ket{\Phi_2(r)}$ transitions the system to an excited manifold, with an energy increase of $qV$ relative to the lowest-energy subspace $q=0$. Within this excited manifold, the unperturbed Hamiltonian $H_{M=6+q}$ features a non-trivial quantum dynamics with spectrum: 
\begin{equation}\label{sm:excited_subspace}
    H_{M=6+q}\ket{\Phi_{2,n q}}=(qV+E^{(0)}_{2,n q})\ket{\Phi_{2,n q}},
\end{equation}
where $\Phi_{2,nq}$ is the eigenstate $n$ in the subspace of $q$ extra nearest-neighbor occupied sites and $E^{(0)}_{2,n q}$ the corresponding eigenvalues where the superscript refer to the fact that it is computed with respect to the unperturbed Hamiltonian $H_{M=6+q}$.  
Expanding the resolvent $1/(H_{6+q}-E^{(0)}_2)$ in the manifold of eigenstates~\eqref{sm:excited_subspace} we find: 
\begin{equation}\label{sm:boson_tunneling2}
    t_b=\sum_{q=1,2}\sum_{n}\frac{\mel{\Phi_2(r+a_j)}{T_{-q}}{\Phi_{2,n q}}\mel{\Phi_{2,n q}}{T_q}{\Phi_2(r)}}{qV+E^{(0)}_{2,n q}+2E_b},
\end{equation}
generalizing the result presented in the maintext to arbitrary $V/\Delta$.
We conclude that the dispersion of the Cooper pair results from the constrained quantum dynamics generated by $H_{M=6+q}$ within the excited energy manifold. 
\begin{figure}
    \centering
    \includegraphics[width=0.7\linewidth]{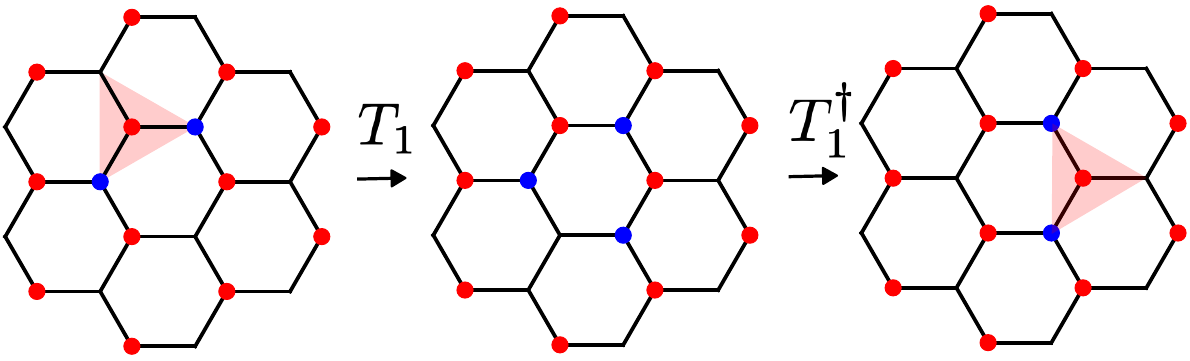}
    \caption{Motion of a Cooper pair from $r$ to $r+a_6$ creating a polaron at $r+a_5$ along the bond $u_1$. The intermediate configuration is connected through $H_0+T_0$ to many other states which are not shown. }
    \label{fig:path1}
\end{figure}
\begin{figure}
    \centering
    \includegraphics[width=1\linewidth]{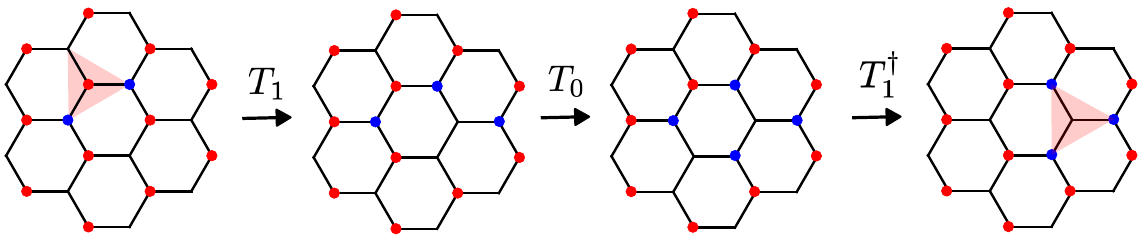}
    \caption{Motion of a Cooper pair from $r$ to $r+a_6$ creating a polaron at $r+a_6$ along the bond $u_1$. The intermediate action of $T_0$ is required in order to connect with the final state.}
    \label{fig:path2}
\end{figure}
We now fix a pair of sites $r$ and $r+a_6$, our task is to determine all the possible second order processes connecting two Cooper pairs. By listing these processes we find that to second order only an intermediate polaron $T_1$ connects the two configurations. In general, there are many independent paths connecting the initial and final configurations, two of them are displayed in Fig~.\ref{fig:path1} and in Fig.~\ref{fig:path2}. The first one connects $\ket{\Phi_2(r)}$ to $\ket{\Phi_2(r+a_6)}$ exciting a polaron along the bond $r+a_5$ to $r+a_{5}+u_1$. The second one, instead, is characterized by the formation of a polaron along the bond $r+a_6$ to $r+a_{6}+u_1$ and requires the intermediate action of $T_{q=0}$ leading to an hopping which does not change the number of n.n. occupied sites. 
\begin{figure}
    \centering
    \includegraphics[width=.7\linewidth]{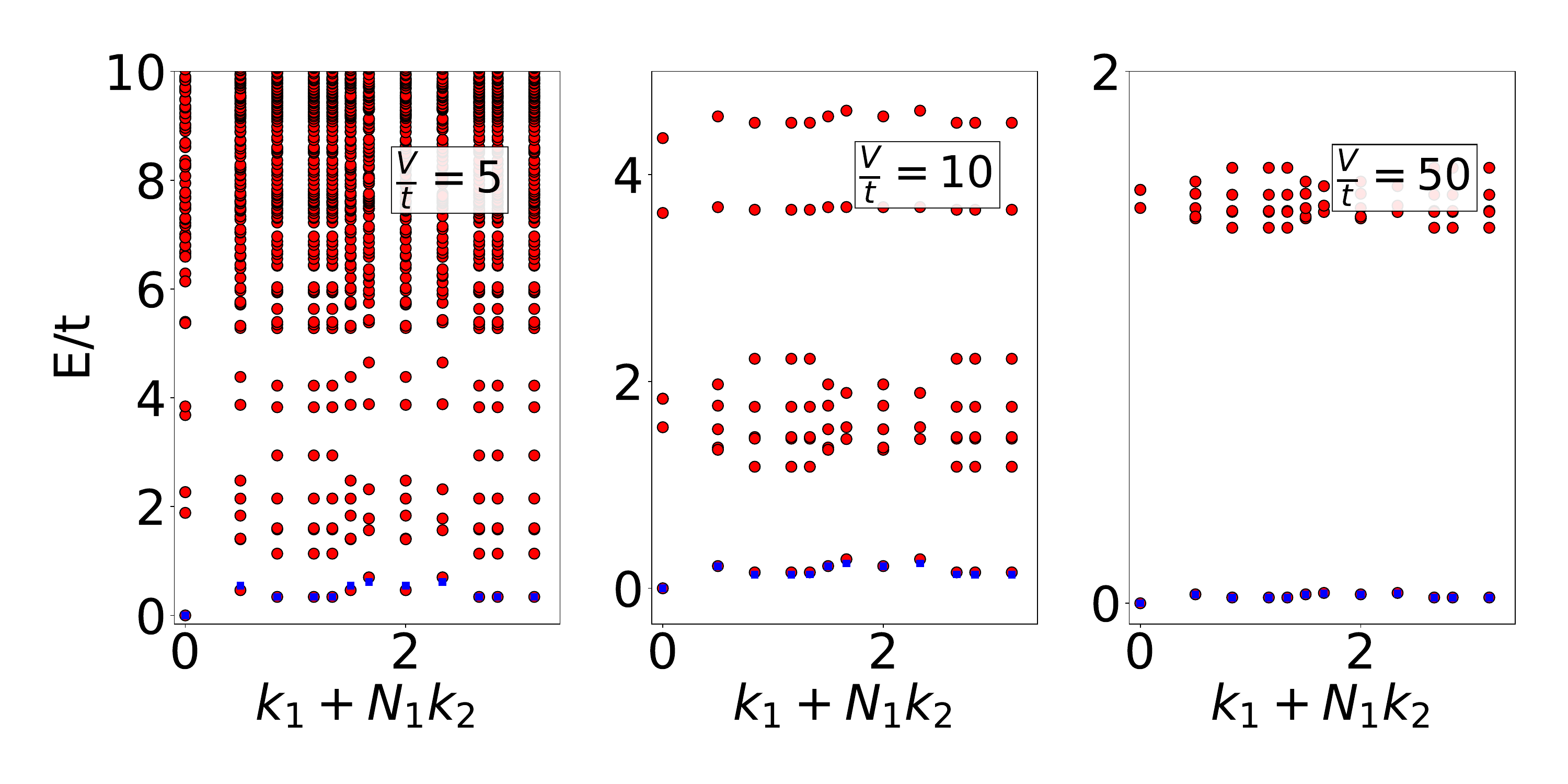}
    \caption{Many-body spectrum for the charge sector with $N_s+2$ particles. The lowest energy band of excitations describes the $2e$ exciton Cooper pair. Blue square data shows the analytical prediction $u_b({\bf q})=-2t_b\sum_{j=1}^{3}\cos({\bf q}\cdot {\bf a}_j)$ for the dispersion relation obtained in perturbation theory. The maximum relative error for the different values of $V/t$ are $17\%$, $12\%$ and $2\%$ for $V/t=5,10,50$, respectively.  We employed $\Delta/t=0.5$. }
    \label{fig:dispersion_2e}
\end{figure}
\begin{figure}
    \centering
    \includegraphics[width=0.7\linewidth]{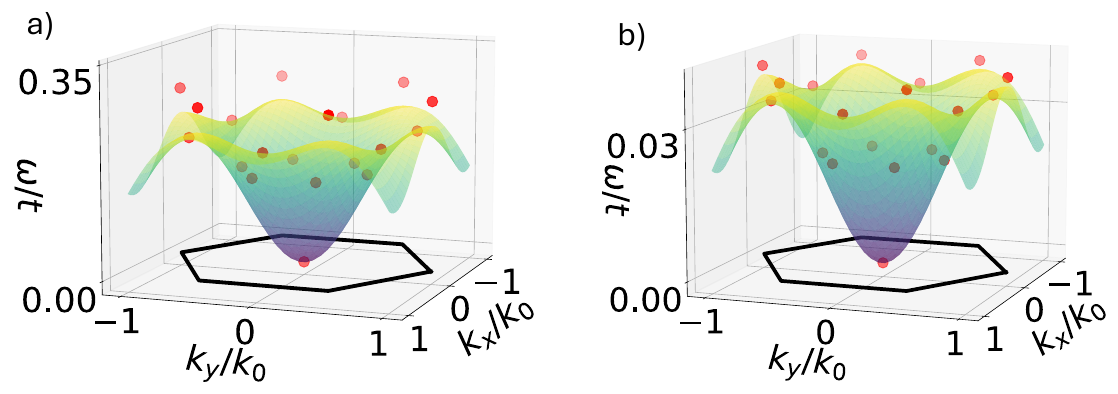}
    \caption{Excitonic Cooper pair dispersion relation for $V/t=10$ (left), $V/t=50$ (right) and $\Delta/t=0.5$. }
    \label{fig:3Ddisperion}
\end{figure}
Including all possible processes we find that the dynamics induced by $H_{M=7}$ is composed by $14$ configurations.

 Fig.~\ref{fig:dispersion_2e} shows the dispersion relation in the sector of $2$ extra doped particles for different values of $V/t$ and $\Delta/t=0.5$. Increasing $V/t$,  our perturbative result offers an increasingly accurate approximation of the dispersion relation. We also show in Fig.~\ref{fig:3Ddisperion} the dispersion relation of the excitonic Cooper pair in the first Brillouin zone. The dispersion relation obtained to first order in $t/V$ accurately captures the energy points across the spectrum, with the exception of the high-symmetry points $K$ and $K'$ at the band edges, where a relative error of $12\%$ for $V/t=10$ is observed.

\subsubsection{charge-$2e$: binding energy}

We now compute the ground state energy gain originating from background intersublattice fluctuations in the presence of the excitonic Cooper pair. In the $V\to\infty$ limit, the ground state energy in this sector is $E^{(0)}_{2}=2\Delta+6V+E_A$ with $E_A$ energy gain in forming the $2e$ excitonic Cooper pair. We perform perturbation theory starting from the ground state with many-body momentum $\Gamma$: 
\begin{equation}\label{gamma_cooper_pair}
    \ket{\Phi_2} = \sum_{r\in A} b^\dagger_r\ket{\Phi_0}/\sqrt{N},
\end{equation}
where $b_r$ is a bosonic operator creating an excitonic Cooper pair at $r$. To compute the first order energy correction we introduce the set of sites $\mathcal S= \{r,r+a_{j=1,\cdots,6}\}$ nearest-neighbors of the Cooper pair centered at $r$. 
The first order correction to the ground state splits in two contributions. The first one involves virtual processes on sites $r\notin\mathcal S$ giving rise to the ground state energy correction:
\begin{equation}
    \delta E_{2,1}=-\frac{}{}(N_s-7)\frac{3t^2}{\Delta + 2V},
\end{equation}
where $7$ counts the number of sites in the set $\mathcal S$. 
The second term instead involves $r\in\mathcal S$ around the Cooper pair: 
\begin{equation}
    \delta E_{2,2} = t^2\sum_{q=1,2}\sum_{r',r'',r'''\in \mathcal S}\sum_{jl=1}^{3}\frac{\mel{\Phi_2(r''')}{\mathbb P_{6}f^\dagger_{r''}f_{r''+u_j}\mathbb P_{6+q}}{\Phi_{2,nq}}\mel{\Phi_{2,nq}}{\mathbb P_{6+q}f^\dagger_{r'+u_l}f_{r'}\mathbb P_{6}}{\Phi_2(r)}}{E^{(0)}_0-E^{(0)}_{2,nq}-qV}, 
\end{equation}
where $\mathbb P_6$ projects in the subspace with $6$ n.n. occupied sites defining the lowest energy subspace for two doped charges while $\mathbb P_{6+q}$ projects in the subspace with $q=1,2$ extra $\sum_{\langle r,r'\rangle} n_{r}n_{r'}$. 
Thanks to the three-fold rotational symmetry and the $C_{2x}$ symmetry we compute the contribution from only one, say $r+a_5$, of the six neighbors in the set $\mathcal S$. 

Several cases arise: the first one (A) involves the hopping operator $-tf^\dagger_{r+a_5+u_2}f_{r+a_5}$ which acts on the ground state $\ket{r}$ only when the configuration  $f^\dagger_{r+u_1}f^\dagger_{r+u_2}\prod_{r'\in A}f^\dagger_{r'}\ket{0}$ is present, weighted by $\alpha/\sqrt{3}$. The configuration $f^\dagger_{r+u_1}f^\dagger_{r+u_2}\prod_{r'\in A}f^\dagger_{r'}\ket{0}$ also couples to an excited state involving a dipole through the action of $-tf^\dagger_{r+a_5+u_{1/3}}f_{r+a_5}$. Summing over these different contributions we find the energy correction 
\begin{equation}
    \delta E_{2,2A} =-6\frac{t^2}{\Delta+2V+2E_b}\alpha^2.
\end{equation}
The second case (B) consists of creating an intermediate polaron configuration which then due to the quantum dynamics introduced by $H_0+T_0$ can either reconnect to the same site or tunnel the Cooper pair to a different site. The latter contribution gives an energy gain $-6t_b$ corresponding to the kinetic energy of a Cooper pair at $Q=\Gamma$.  The energy variation reads: 
\begin{equation}
    \delta E_{2,2B} = -6t_b + \sum_{r',r''\in \mathcal S}\sum_{jl=1}^{3}\frac{\mel{\Phi_2(r)}{\mathbb P_{6}f^\dagger_{r''}f_{r''+u_j}\mathbb P_{7}}{\Phi_{2,n1}}\mel{\Phi_{2,n1}}{\mathbb P_{7}f^\dagger_{r'+u_l}f_{r'}\mathbb P_{6}}{\Phi_2(r)}}{E^{(0)}_0-E^{(0)}_{\gamma,1}-V}.
\end{equation}
The resulting second-order energy correction is given by the sum $\delta E_{2}=\delta E_{2,1}+\delta E_{2,2A}+\delta E_{2,2B}$. 

\begin{figure}
    \centering
    \includegraphics[width=0.6\linewidth]{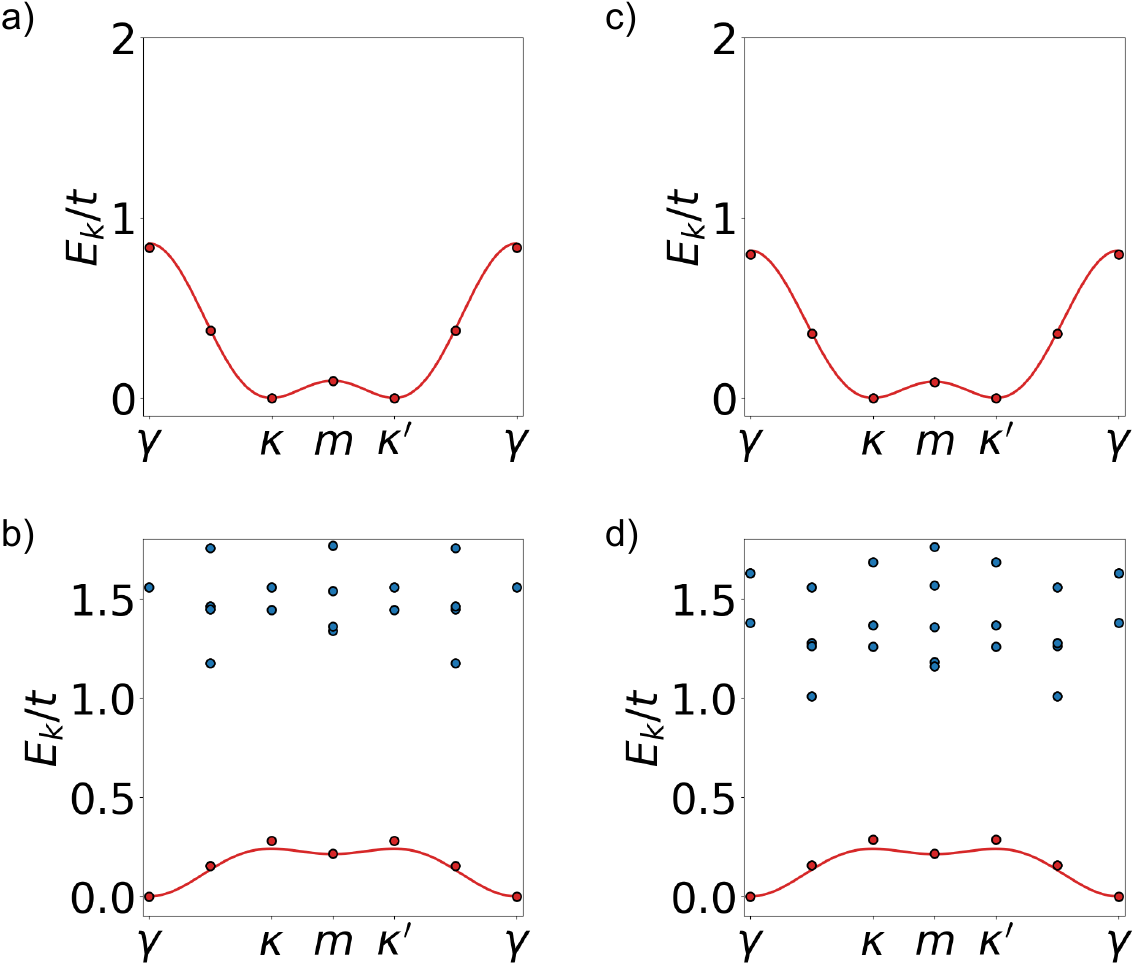}
    \caption{Panels a)-c) and b)-d) shows the dispersion relation of charge-$e$ and $2e$ excitations, respectively, for $V/t=10$ and $\Delta/t=0.5$ (left column) and $\Delta/t=1$ (right column).}
    \label{fig:high_symmetry_dispersion}
\end{figure}

\section{Low-energy effective field theories}
\label{app:field_theory}

In this section, we present analytical results obtained by employing field theoretical approach valid in the ``ionic'' regime of $\Delta\gg t$.

\subsection{Doping the charge transfer insulator $\Delta/t\gg 1$}

\begin{figure}
    \centering
    \includegraphics[width=.8\linewidth]{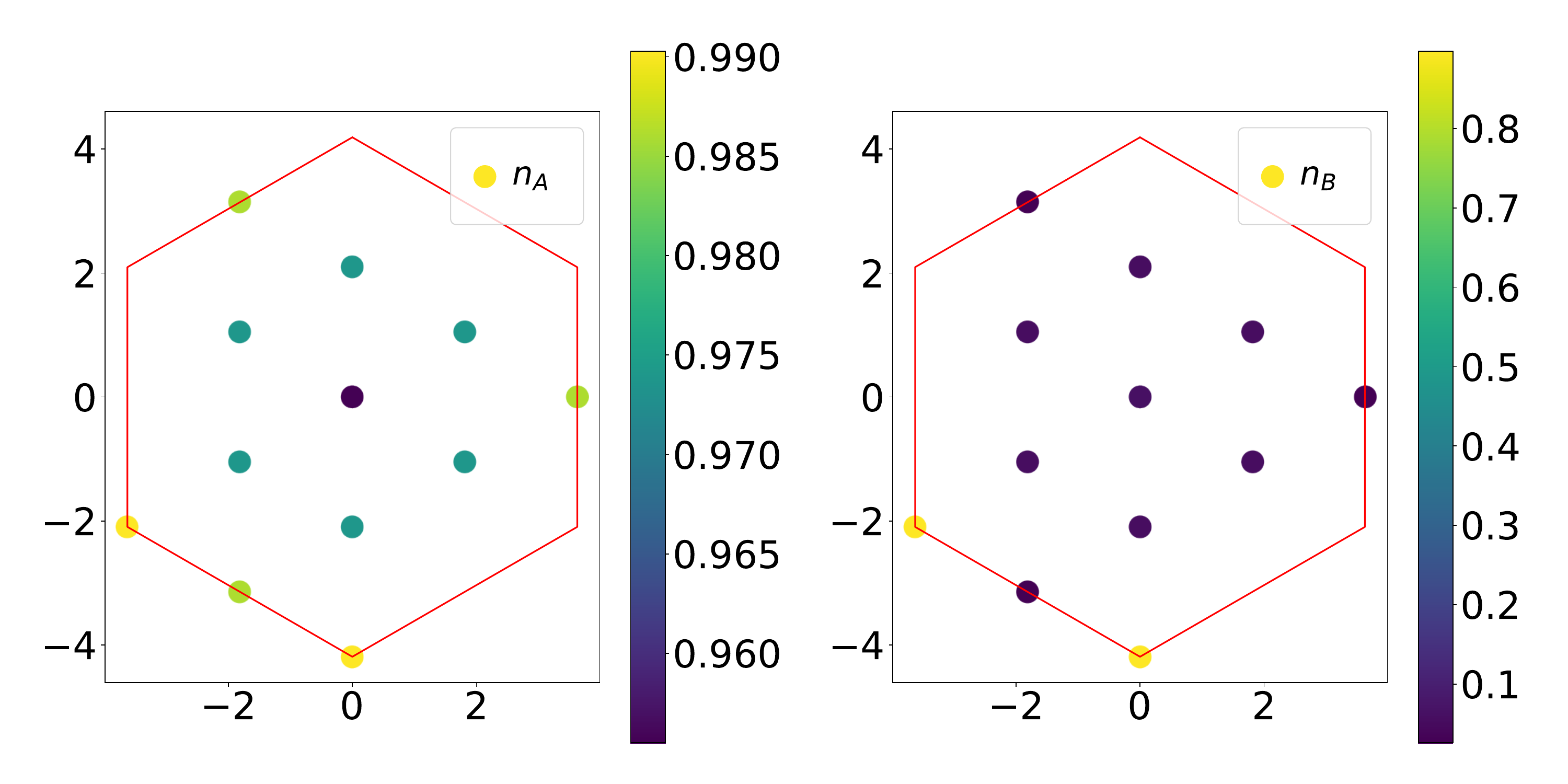}
    \caption{Momentum space distribution for the two different sublattices for $N_p=14$ (filling factor $1/2+1/12$) also equal to $1/6$ in the $B$ sublattice, $V/t=10$ and $\Delta/t=3$. 
    The sum over momenta gives $\sum_{k} \rho_{k,AA}/N_s\approx0.978$ and $\sum_{k} \rho_{k,BB}/N_s\approx0.189$ ($1/6=0.167$ smaller due to intersublattice fluctuations). The color code shows $\langle n_{kA/B} \rangle$.}
    \label{fig:momentum_space_2e}
\end{figure}

In the long wavelength limit, doped carriers are located around $K$ and $K'$ (see Fig.~\ref{fig:momentum_space_2e}) denoted as $\pm$ and their motion is described by: 
\begin{equation}\label{sm:continuum}
    \mathcal H = \int \frac{d^2 x}{\Omega}\sum_\tau \psi_\tau^\dagger \left(-\frac{\nabla^2_r}{2m_f}\right) \psi_\tau - g\psi^\dagger_+\psi^\dagger_-\psi_-\psi_+,
\end{equation}
where $\psi_+/\psi_-$ are Fermi fields for valley $K/K'$, respectively, $m_f=2/(3t_f)$, $t_f=t^2/(\Delta+V)$ and $g=6(2\lambda-V_f)=36t^2V^2/[\Delta(\Delta+V)(\Delta+2V)]>0$ attractive for arbitrary $V$ and $\Delta$~\cite{Cr_pel_2022}. 
The problem of two doped carriers reduces to the solution of the Schr\"odinger equation for the two particle bound state: 
\begin{equation}\label{2e_bound_state_continuum}
    \ket{\Psi_{2e}} = \int_p F(p) \psi^\dagger_{p+}\psi^\dagger_{-p-}\ket{0},
\end{equation}
where the two particle wavefunction $F(p)=F(-p)$ is given by: 
\begin{equation}
    F(p) = \frac{g}{E_b+p^2/2m_f},
\end{equation}
with $E_b$ binding energy per particle. 
The binding energy is obtained solving the equation: 
\begin{equation}
    \frac{1}{g}=\frac{1}{2}\int^\Lambda \frac{d^2p}{4\pi^2}\frac{1}{|E_b|+p^2/2m_f}\implies  E_b = \frac{\epsilon_\Lambda}{e^{1/\lambda}-1},\quad \lambda=\frac{|g|m_f}{4\pi}.
\end{equation}
Here, $\epsilon_\Lambda=\Lambda^2/(2m_f)$ is the ultraviolet cutoff, chosen such that $E_b$ agrees with the lattice result in the regime $V\gg\Delta$ and $\Delta\gg t$. Employing Eq.~(2) of the main text, we find $E_b\approx 3t^2/(2\Delta)$ which implies $\Lambda=\sqrt{2\pi/3}$ $(a=1)$.

The Cooper pair wavefunction is given by: 
\begin{equation}
    \Psi_{2e}(\Delta r) = \mel{0}{\psi_-(r+\Delta r)\psi_{+}(r)}{\Psi_{2e}}=\int_k F(k) e^{ik\cdot\Delta r}.
\end{equation}
 The resulting mean square radius reads: 
\begin{equation}
    \langle r^2\rangle = \frac{\int_r  r^2 |\Psi_{2e}(r)|^2}{\int_r  |\Psi_{2e}(r)|^2}=\frac{\int_k |\nabla_k F(k)|^2}{\int_k |F(k)|^2}=\frac{1}{3m_f E_b},
\end{equation}
where the last relation follows from Parseval theorem. Finally, the dispersion of the Cooper pairs is obtained expanding the energy of the bound state~\eqref{2e_bound_state_continuum} for finite center of mass momentum and reads: 
 \begin{equation}
     \frac{1}{m_b}=\frac{1}{2m_f}=\frac{3t^2}{4(\Delta +V)},
 \end{equation}
 following the same asymptotic behavior as the result obtained in the main text but lacking the correct prefactors.

\end{document}